\documentclass[12pt,manuscript]{aastex}
\usepackage[graphicx]{}
\usepackage{multirow}

\newcommand{\vy}[2]{#1_{\scriptscriptstyle #2}}

\def\gtorder{\mathrel{\raise.3ex\hbox{$>$}\mkern-14mu
             \lower0.6ex\hbox{$\sim$}}}
\def\ltorder{\mathrel{\raise.3ex\hbox{$<$}\mkern-14mu
             \lower0.6ex\hbox{$\sim$}}}
\def\proptwid{\mathrel{\raise.3ex\hbox{$\propto$}\mkern-14mu
             \lower0.6ex\hbox{$\sim$}}}
\def\0946{PG~0946+301}

\def\arcsec{\ifmmode '' \else $''$\fi}

\def\arcsecpoint{\ifmmode ''\!. \else $''\!.$\fi}

\def\kms{\ifmmode {\rm km\ s}^{-1} \else km s$^{-1}$\fi}
\def\Msun{\ifmmode {\rm M}_{\odot} \else M$_{\odot}$\fi}
\def\Lsun{\ifmmode {\rm L}_{\odot} \else L$_{\odot}$\fi}
\def\Zsun{\ifmmode {\rm Z}_{\odot} \else Z$_{\odot}$\fi}

\def\ergscm2{ergs\,s$^{-1}$\,cm$^{-2}$}
\def\icm3{{\rm cm}^{-3}}
\def\icm2{{\rm cm}^{-2}}
\def\qo{\ifmmode q_{\rm o} \else $q_{\rm o}$\fi}
\def\Ho{\ifmmode H_{\rm o} \else $H_{\rm o}$\fi}
\def\ho{\ifmmode h_{\rm o} \else $h_{\rm o}$\fi}
\def\ltsim{\raisebox{-.5ex}{$\;\stackrel{<}{\sim}\;$}}

\def\vFWHM{\ifmmode v_{\mbox{\tiny FWHM}} \else
            $v_{\mbox{\tiny FWHM}}$\fi}
\def\CCF{\ifmmode F_{\it CCF} \else $F_{\it CCF}$\fi}
\def\ACF{\ifmmode F_{\it ACF} \else $F_{\it ACF}$\fi}
\def\Halpha{\ifmmode {\rm H}\alpha \else H$\alpha$\fi}
\def\Hbeta{\ifmmode {\rm H}\beta \else H$\beta$\fi}
\def\Hgamma{\ifmmode {\rm H}\gamma \else H$\gamma$\fi}
\def\Hdelta{\ifmmode {\rm H}\delta \else H$\delta$\fi}
\def\Lya{\ifmmode {\rm Ly}\alpha \else Ly$\alpha$\fi}
\def\Lyb{\ifmmode {\rm Ly}\beta \else Ly$\beta$\fi}
\def\Lyg{\ifmmode {\rm Ly}\beta \else Ly$\gamma$\fi}
\def\hi{H\,{\sc i}}

\def\hei{He\,{\sc i}}
\def\heii{He\,{\sc ii}}

\def\cii{C\,{\sc ii}}
\def\ciii{\ifmmode {\rm C}\,{\sc iii} \else C\,{\sc iii}\fi}
\def\civ{\ifmmode {\rm C}\,{\sc iv} \else C\,{\sc iv}\fi}

\def\nv{N\,{\sc v}}

\def\oii{O\,{\sc ii}}

\def\o5007{[O\,{\sc iii}]\,$\lambda5007$}
\def\oiv{O\,{\sc iv}}

\def\ovi{O\,{\sc vi}}

\def\neviii{Ne\,{\sc viii}}

\def\mgii{Mg\,{\sc ii}}
\def\mgx{Mg\,{\sc x}}
\def\siiv{Si\,{\sc iv}}

\def\siIII{Si\,{\sc iii}}
\def\siII{Si\,{\sc ii}}

\def\siv{S\,{\sc iv}}

\def\feii{Fe\,{\sc ii}}
\def\feiii{Fe\,{\sc iii}}

\def\alii{Al\,{\sc ii}}
\def\aliii{Al\,{\sc iii}}

\def\pv{P\,{\sc v}}

\def\o{\o}
\begin{document}

\title{Major contributor to AGN feedback:\\
VLT X-shooter observations of \siv\ BAL QSO outflows\footnote{Based on observations collected at the European Southern Observatory, Chile, PID:87.B-0229}}

\author{
 Benoit C.J. Borguet\altaffilmark{1},
 Nahum Arav\altaffilmark{1},
 Doug Edmonds\altaffilmark{1},
 Carter Chamberlain\altaffilmark{1}, 
 Chris Benn\altaffilmark{2}
}

\date{Version : \today}

\altaffiltext{1}{Department of Physics, Virginia Tech, Blacksburg, VA 24061; email: benbo@vt.edu}
\altaffiltext{2}{Isaac Newton Group, Apartado 321, 38700 Santa Cruz de La Palma, Spain}

\begin{abstract}

We present the most energetic BALQSO outflow measured to date, with a
kinetic luminosity of at least $10^{46}$ ergs s$^{-1}$, which is 5\% of the
bolometric luminosity of this high Eddington ratio quasar. The associated
mass flow rate is 400 solar masses per year. Such kinetic luminosity
and mass flow rate should provide strong AGN feedback effects.
The outflow is located at about 300 pc from the quasar and has a
velocity of roughly 8000 km s$^{-1}$. Our distance and energetic measurements
are based in large part on the identification and measurement of \siv\
and \siv* BALs.  The use of this high ionization species allows us to generalize
the result to the majority of high ionization BALQSOs that are identified
by their \civ\ absorption. We also report the energetics of two other
outflows seen in another object using the same technique. The distances
of all 3 outflows from the central source (100-2000pc) suggest that we
observe BAL troughs much farther away from the central source than the assumed acceleration region
of these outflows (0.01-0.1pc).

\end{abstract}


\section{INTRODUCTION}
\label{intro}

Broad absorption line (BAL) outflows are observed as blueshifted troughs in the
rest-frame spectrum of $\sim$20 \% of quasars
\citep{Hewett03,Ganguly08,Knigge08}.
The energy, mass, and momentum carried by these outflows are thought to play a
crucial role in shaping the early universe and dictating its evolution
\citep[e.g.][]{Scannapieco04,Levine05,Hopkins06,Cattaneo09,Ciotti09,Ciotti10,Ostriker10}.
The ubiquity and wide opening angle deduced from the detection rate of
these mass
outflows, allows for efficient interaction with the surrounding medium,
and these outflows carry thousands of times more mass flux per unit of
kinetic luminosity than
the collimated relativistic jets observed in 5 -- 10\% of all AGNs.
Theoretical studies and simulations show that this so-called AGN feedback can provide
an explanation for a variety of observations, from the chemical enrichment
of the intergalactic medium, to the self regulation of the growth of the
supermassive black-hole and of the galactic bulge \citep[e.g.][and references
therein]{Silk98,dimatteo05,Germain09,Hopkins09,Elvis06}.

Quantifying AGN feedback requires estimating the kinetic luminosity
($\dot{E}_k$)
and mass-flow rate ($\dot{M}$) of the outflows. These quantities can be computed
in cases where we are able to estimate the distance $R$ to the outflowing
material from the central source (see Equations 6 and 7 in \citealt{Borguet12a}).
Following the definition of the ionization parameter of the plasma $U_H \propto
1/\vy{n}{H} R^2$ where $\vy{n}{H}$ is the hydrogen number density, the distance
can be obtained for outflows of known ionization state and density (see elaboration in Section~\ref{firstorder}). The
research program developed by our team has led to the
determination
of $\dot{M}$ and $\dot{E}_k$ in several quasar outflows
\citep[e.g.][]{Moe09,Dunn10a,Bautista10,Aoki11,Borguet12a} by using the population
ratio of collisionaly
excited states to the resonance levels of singly ionized species (e.g. \siII,
\feii) as density diagnostics \citep[see][for a review]{Crenshaw03}. The lower
detection
rate of these low ionization outflows in spectroscopic surveys raises the
question of whether the determinations obtained for these objects are
representative
of the ubiquitous high ionization \civ\ broad absorption line quasars
\citep[see][hereafter Paper I]{Dunn12}.

One way to alleviate this uncertainty is to target objects which possess
absorption troughs from excited states of high ionization species, where \siv/\siv*
$\lambda\lambda 1062.66,1072.97$ are especially promising. These transitions appear at wavelengths
blueward of Ly$\alpha$ and therefore suffer from blending with the
Lyman forest in high redshift objects. However, since the ratio of
\civ\ and \siv\ ionic
fractions as a function of the ionization parameter ($U_H$)
is relatively constant (see paper I) they must arise from the same
photoionized plasma.
The ionization similarity of \civ\ and \siv\ makes \siv/\siv* outflows
a much better agent than the usual low ionization species for the
determination of the feedback from the high ionization
outflows (see discussion in Section~\ref{omegadisku}). We consequently developed a
research program that measured the sample properties of these
\siv\ outflows (Paper I) and analyzed the photoionization and chemical
abundances of one such outflow (Borguet et al. 2012b, hereafter paper II).
In this paper, we present the analysis of VLT/X-shooter spectra of two SDSS BAL
quasars, SDSS J1106+1939 and SDSS J1512+1119, for which the presence of
\siv/\siv* troughs allows us to place constraints on the location and energetics of the outflow
in SDSS J1106+1939 and of two separate outflows in SDSS J1512+1119.

The plan of the paper is as follows: In \S~\ref{dataredu} we present the
VLT/X-shooter observations of SDSS J1106+1939 and SDSS J1512+1119 along with
the reduction of the data. In \S~\ref{anabs} we identify the spectral features
associated with \siv\ and \siv* and measure the column densities for
various ionic species associated with each outflow. Photoionization models
are crucial for finding the total hydrogen column density of the
outflows and their ionization equilibrium.
We present these models in \S~4.
In \S~\ref{firstorder} we derive
the parameters  necessary to determine $R$,
$\dot{E}_k$ and $\dot{M}$ of the \siv\ outflows.
In \S~\ref{cav}, we evaluate the robustness of each step in the analysis and conclude that $\dot{E}_k=10^{46}$
ergs~s$^{-1}$ is a firm lower limit. We discuss our results in \S~\ref{discufin}.

\section{OBSERVATIONS AND DATA REDUCTION}
\label{dataredu}

We selected our targets SDSS J1106+1939 (J2000: RA=11 06 45.05, dec=+19 39 29.1, g=19.4) and SDSS J1512+1119 (J2000: RA=15 12 49.29, dec=+11 19 29.4, g=17.7)
from an extensive flux limited search of the SDSS DR7 catalog for objects indicative of the presence of \siv\ absorption on the basis of their SDSS spectrum.
We observed these two objects with VLT/X-shooter as part of our program 87.B-0229 (PI: Benn) in April 2011 and March 2012, respectively. X-shooter is the second generation,
medium spectral resolution ($R \sim$ 6000--9000) spectrograph installed at the Cassegrain focus of VLT/UT2 \citep{vernet11}. The unique design of the instrument, in which the incoming light is split into three independent
arms (UVB, VIS and NIR) each composed of a prism-cross-dispersed echelle spectrograph, allows the simultaneous covering of a wide spectral band (3000 \AA ~to 24000 \AA)
in a single exposure. The total integration time for each object in the UVB, VIS and NIR arms were 8400, 8400, and 8700 s, respectively.

\begin{figure}
  \includegraphics[angle=90,width=1.0\textwidth]{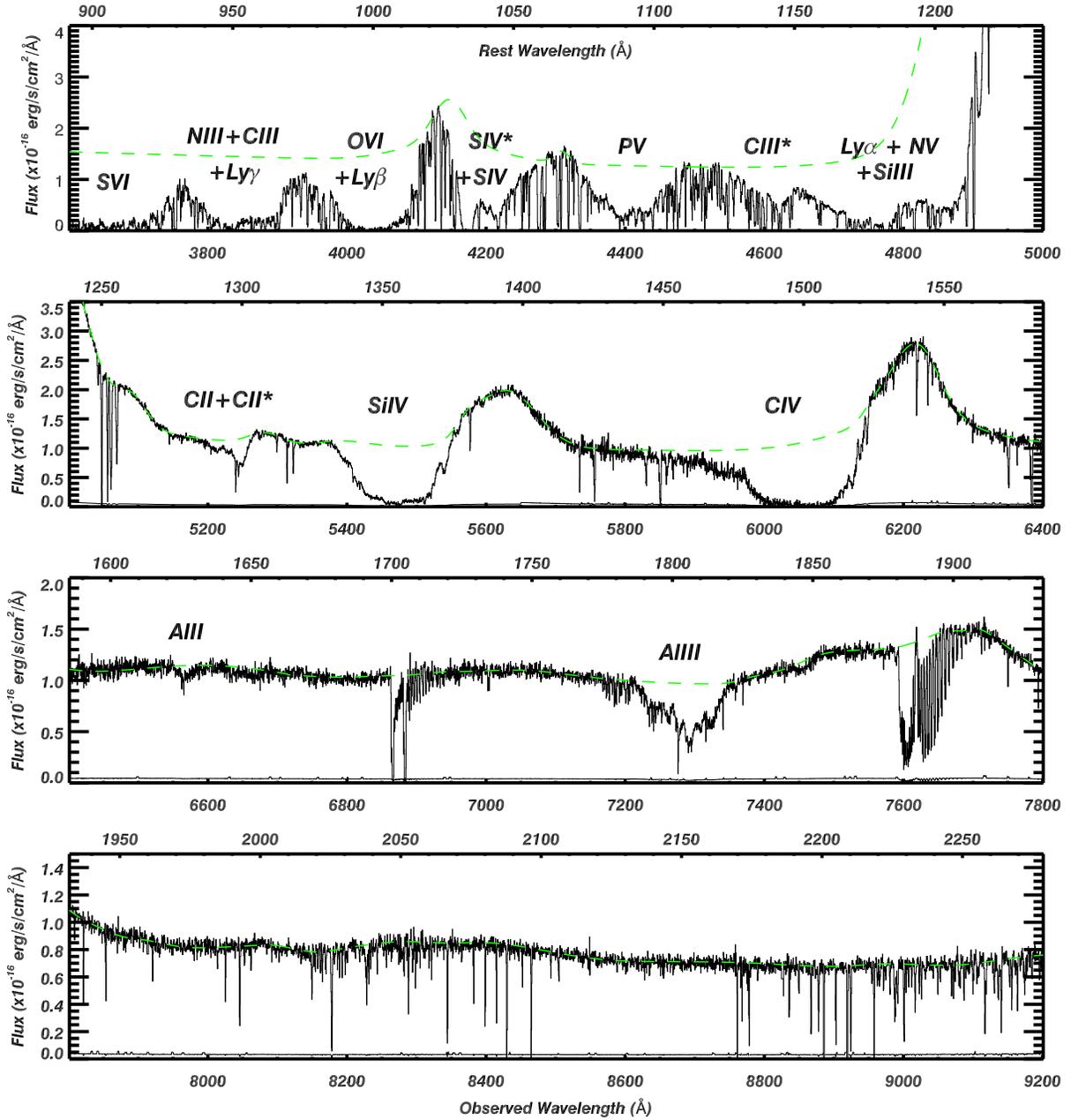}\\
 \caption{Reduced and fluxed UVB+VIS X-shooter spectra of the quasar SDSS J1106+1939. We
indicate the position of the absorption lines associated with the intrinsic outflow.
The VIS part of the spectrum ($\lambda_{obs} > 5700$ \AA) has not been corrected for atmospheric
absorption. This does not affect our study since our diagnostic lines are located in regions
free of such contamination. The dashed line represents our unabsorbed emission model (see Section~\ref{unabmod}).}
 \label{fullspec1106} 
\end{figure}

We reduced the SDSS J1106+1939 spectra in an identical fashion to the one of SDSS J1512+1119 (detailed in Paper II): we rectified and wavelength calibrated
the two dimensional spectra using the ESO Reflex workflow \citep{ballester11}, then extracted one dimensional spectra using an optimal extraction algorithm
and finally flux calibrated the resulting data with the spectroscopic observations of a standard star observed the same day as the quasar. The resulting flux
calibrated spectrum of SDSS J1106+1939 is presented in Figure~\ref{fullspec1106}. We only present the UVB+VIS spectrum and present the few additional diagnostic
absorption lines detected in the NIR portion of the spectrum in Fig.~\ref{figfirstshot}.

\section{SPECTRAL FITTING}
\label{anabs}

The radial velocity values across the absorption troughs are determined with respect to the systemic redshift of the quasar. 
\citet{Hewett10} report an improved redshift value\footnote{Available online at: http://das.sdss.org/va/Hewett\_Wild\_dr7qso\_newz/}
for these two SDSS DR7 quasars by cross correlating a quasar emission line template to the observed SDSS spectrum. These improved
redshifts are $z=3.0377 \pm 0.0021$ for SDSS J1106+1939 and  $z=2.1062 \pm 0.0020$ for SDSS J1512+1119.
Note that while the NIR range of the X-shooter observations of J1106+1939 covers the \Hbeta + [O\,{\sc iii}]\,$\lambda5007$ and also the \mgii\ rest frame regions,
these portions of the spectrum are located within dense and strong H$_2$O and CO$_2$ atmospheric bands preventing us
from determining a more accurate redshift from the fit of these emission features. We also examined the expected position of [\oii]\ $\lambda$3727 and
found no emission feature in its spectral vicinity.

%
%
%
%

\subsection{Unabsorbed Emission Model}
\label{unabmod}

Deriving ionic column densities from absorption troughs requires knowledge of the underlying unabsorbed emission $F_0(\lambda)$.
In AGNs the UV unabsorbed emission source can generally be decomposed into two components: a continuum source described by
a power law and emission lines usually modeled by Gaussian profiles that are divided into broad emission lines (BEL, full width at half maximum (FWHM)  $> 1000$ km s$^{-1}$)
and narrow emission lines (NEL, FWHM $< 1000$ km s$^{-1}$).

Due to the presence of strong absorption shortward of the rest-frame Ly$\alpha$ emission in SDSS J1106+1939, we fit the continuum in the range 3200 -- 9000 \AA\
using a single power law of the form $F(\lambda)=F_{1100} (\lambda/1100)^\alpha$, where $F_{1100}$ is the observed flux at 1100 \AA\ in the rest-frame
of the object, on regions longward of the Ly$\alpha$ emission suspected free of absorption or emission lines. After correcting the spectrum for the
galactic extinction (E(B-V) = 0.025, \citealt{Schlegel98}) using the reddening curve of \citet{Cardelli89} we find $F_{1100}=12.8 \times 10^{-17}$
ergs/s/\AA/cm$^{2}$ and $\alpha \simeq -0.95$. We fit the prominent Ly$\alpha$ and \civ\ emission lines using a sum of Gaussians and a spline fit 
for the \siiv +\oiv$]$ blend. We fit the \feii\ and other weak emission lines longward of 1600 \AA\ in the rest frame using a phenomenological spline fit. In the far UV, we build an \ovi\ emission model by scaling and shifting to the proper position the \civ\ emission line model. The scaling factor is chosen
to be the smallest such as no emission remains above the constructed emission model. This approach could strongly underestimate the true underlying \ovi\ emission.
We fit the NIR unabsorbed emission using a smooth spline fit. The resulting emission model for the UVB+VIS region is shown in Figure~\ref{fullspec1106}.
Note that the single power law model for the whole UVB+VIS range most likely provides an overestimation of the true underlying continuum shortward of
the \ovi\ emission, a region in which a softer power law is usually adopted \citep{Korista92,Zheng97,Arav01,Telfer02}. However, modeling of the unabsorbed
emission in that region is not important for our BAL analysis due to the fact that it contains only heavily blended diagnostic lines coupled with a
limited signal to noise ratio (S/N).

Given the overall high S/N of the SDSS J1512+1119 data spectrum (S/N $\sim$ 30 -- 70 over most of the UVB/VIS range) and absence of
wide absorption troughs, we fit the unabsorbed emission lines using a smooth third order spline fit 
after fitting the derredenned continuum (E(B-V) = 0.051, \citealt{Schlegel98}) with a power law with
$F_{1100}=77.0 \times 10^{-17}$ ergs/s/\AA/cm$^{2}$ and $\alpha \simeq -0.403$ (see Paper II for details).


\subsection{BAL Column Density Measurements of SDSS J1106+1939}
\label{identifeat}

\begin{figure}
  \includegraphics[angle=90,width=0.31\textwidth]{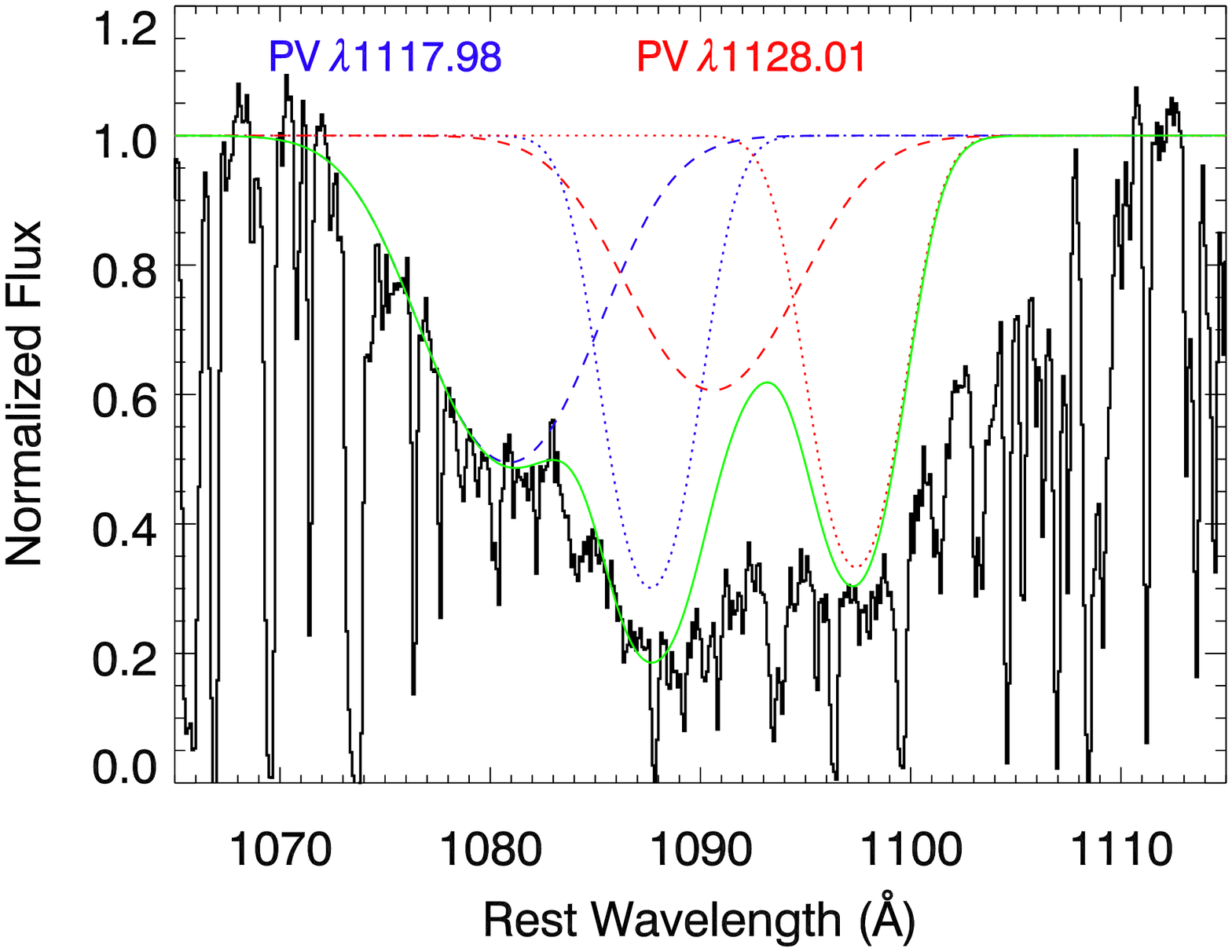}
  \includegraphics[angle=90,width=0.31\textwidth]{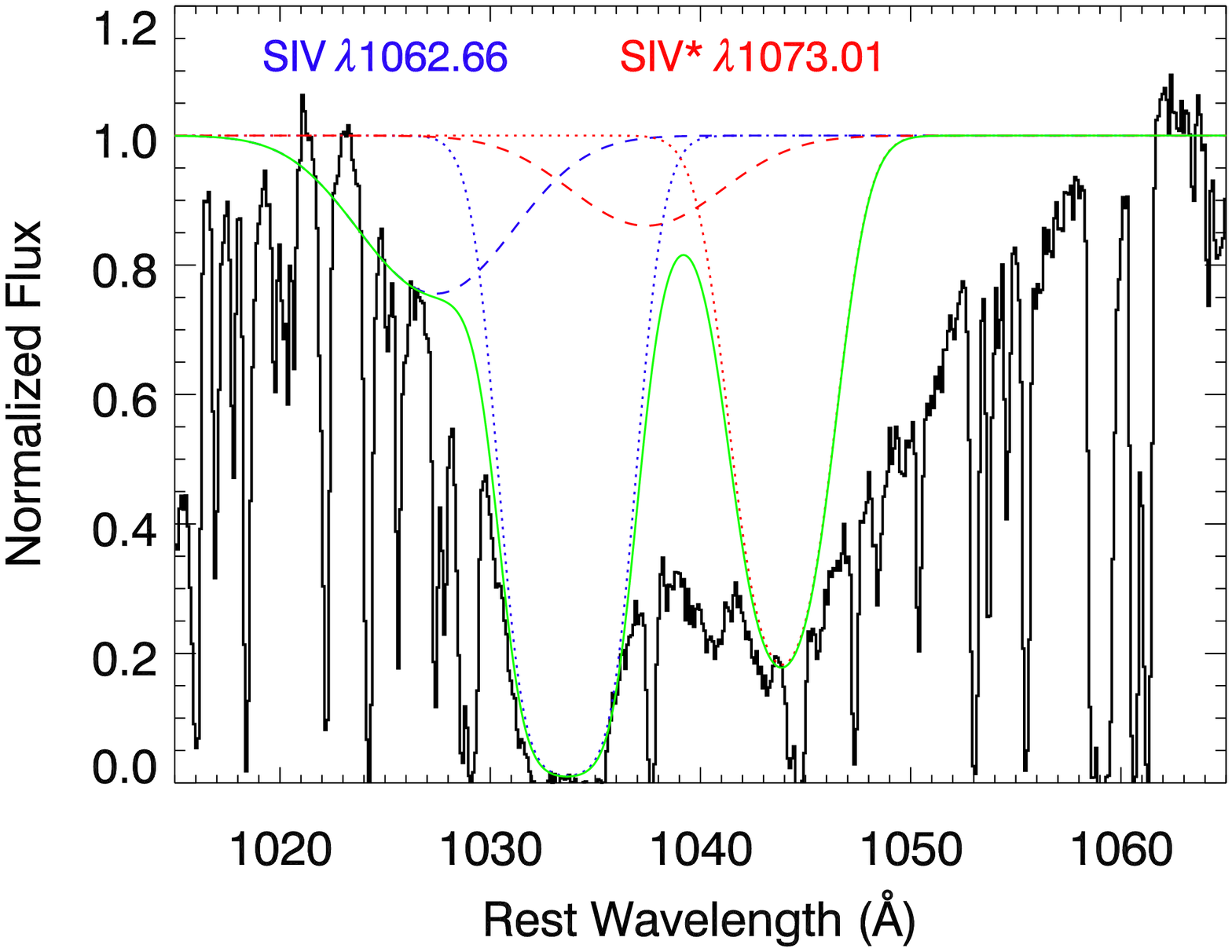}
  \includegraphics[angle=90,width=0.31\textwidth]{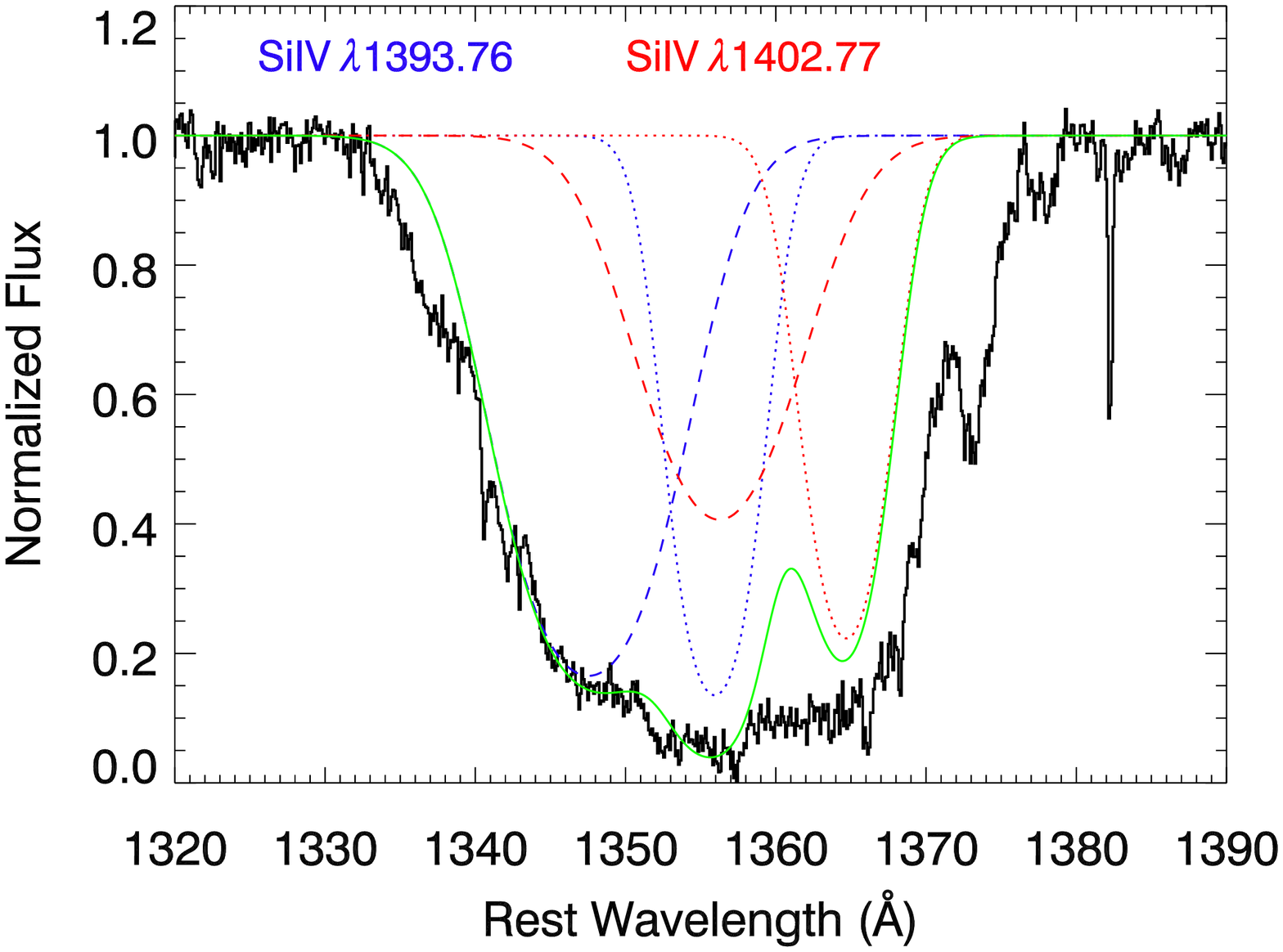}\\
  \includegraphics[angle=90,width=0.31\textwidth]{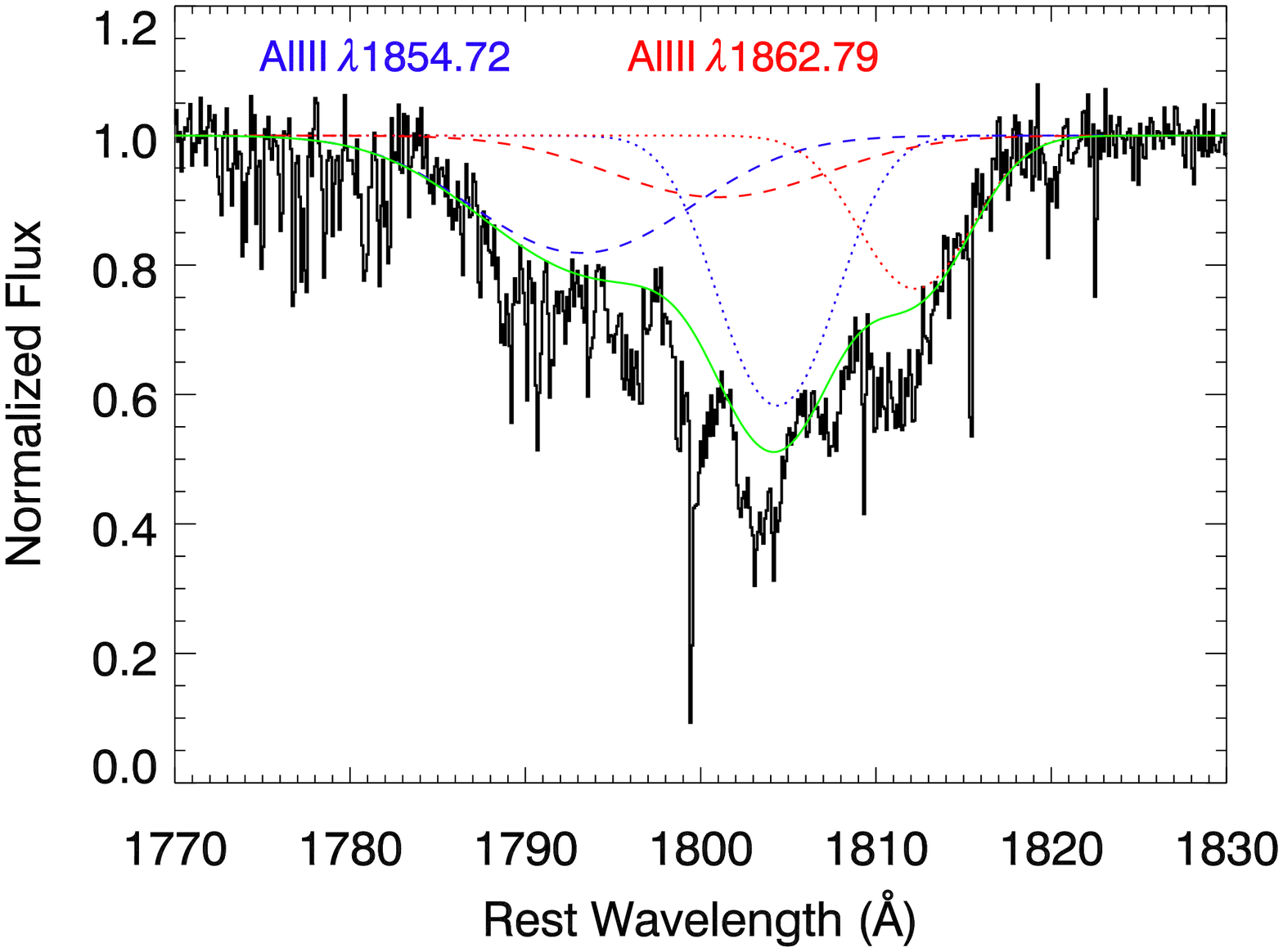}
  \includegraphics[angle=90,width=0.31\textwidth]{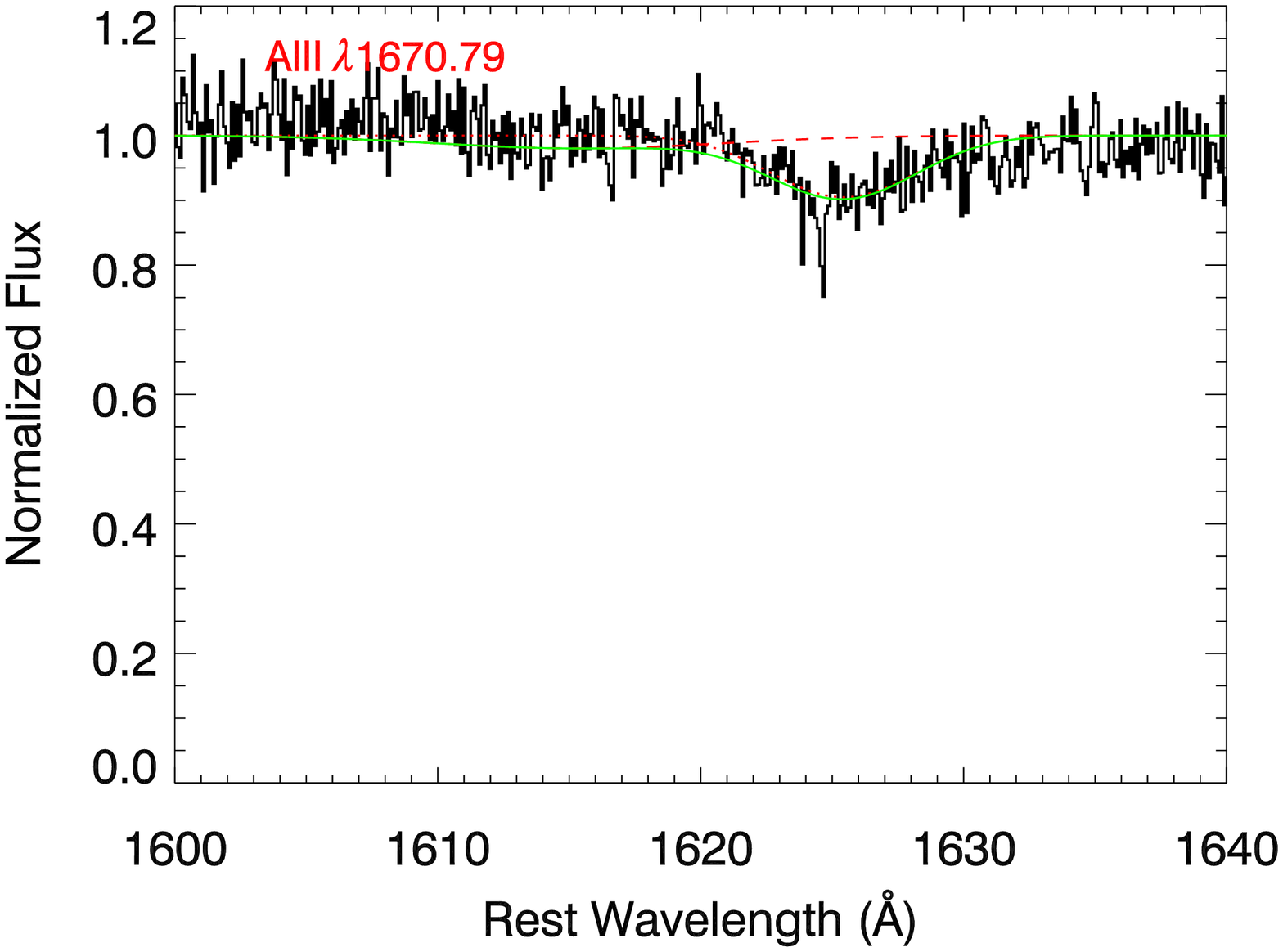}
  \includegraphics[angle=90,width=0.31\textwidth]{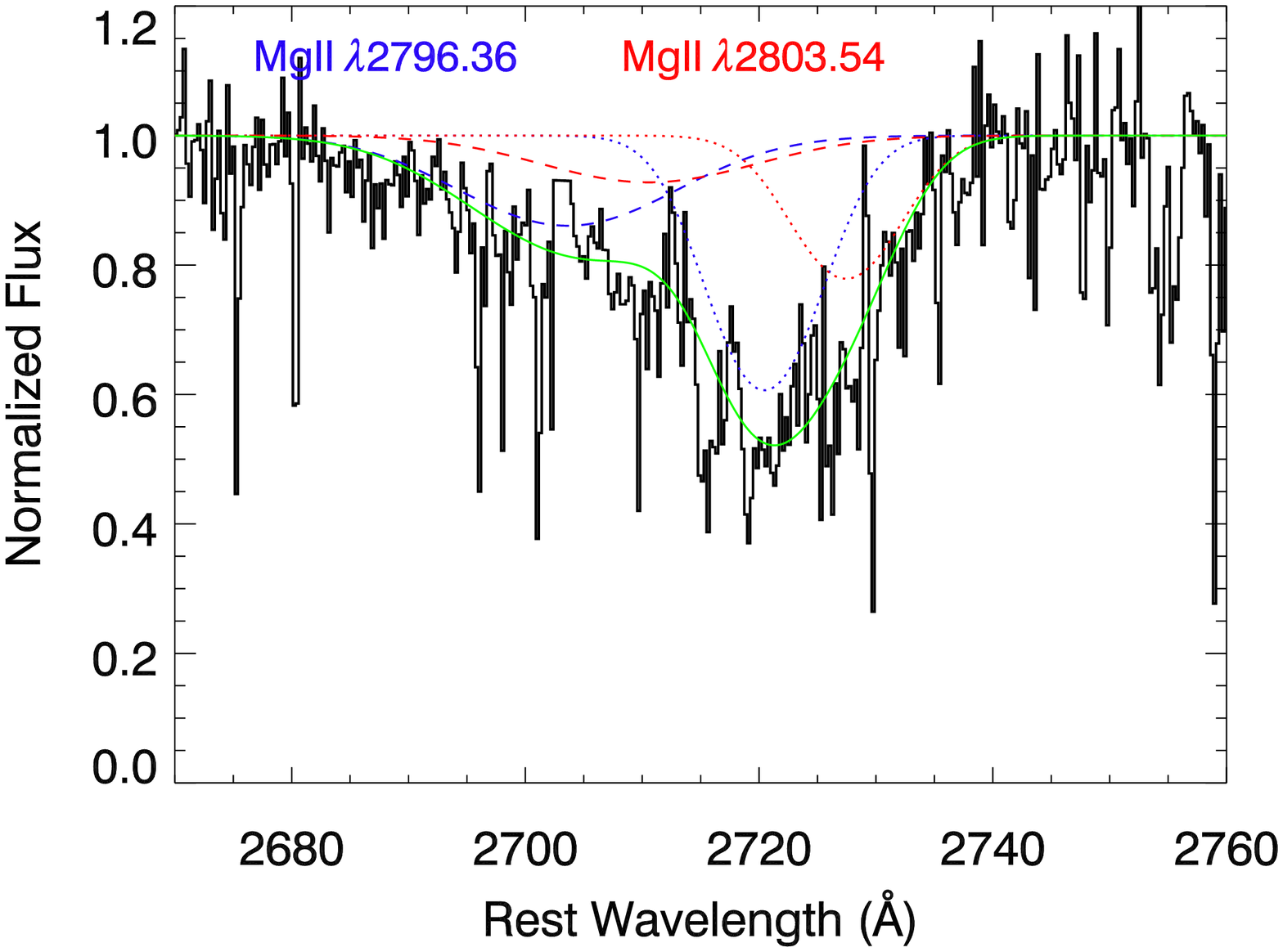}\\
  \includegraphics[angle=90,width=0.31\textwidth]{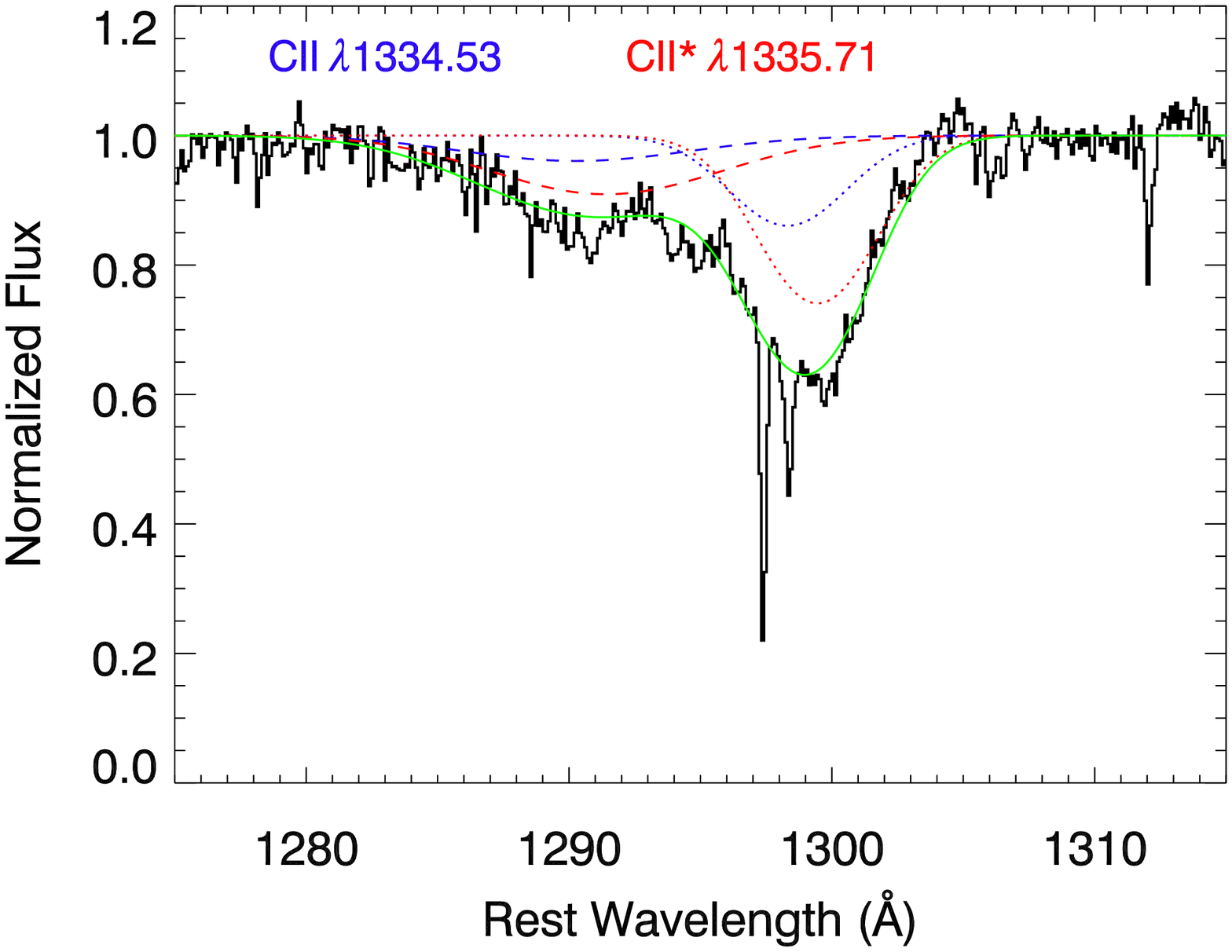}
  \includegraphics[angle=90,width=0.31\textwidth]{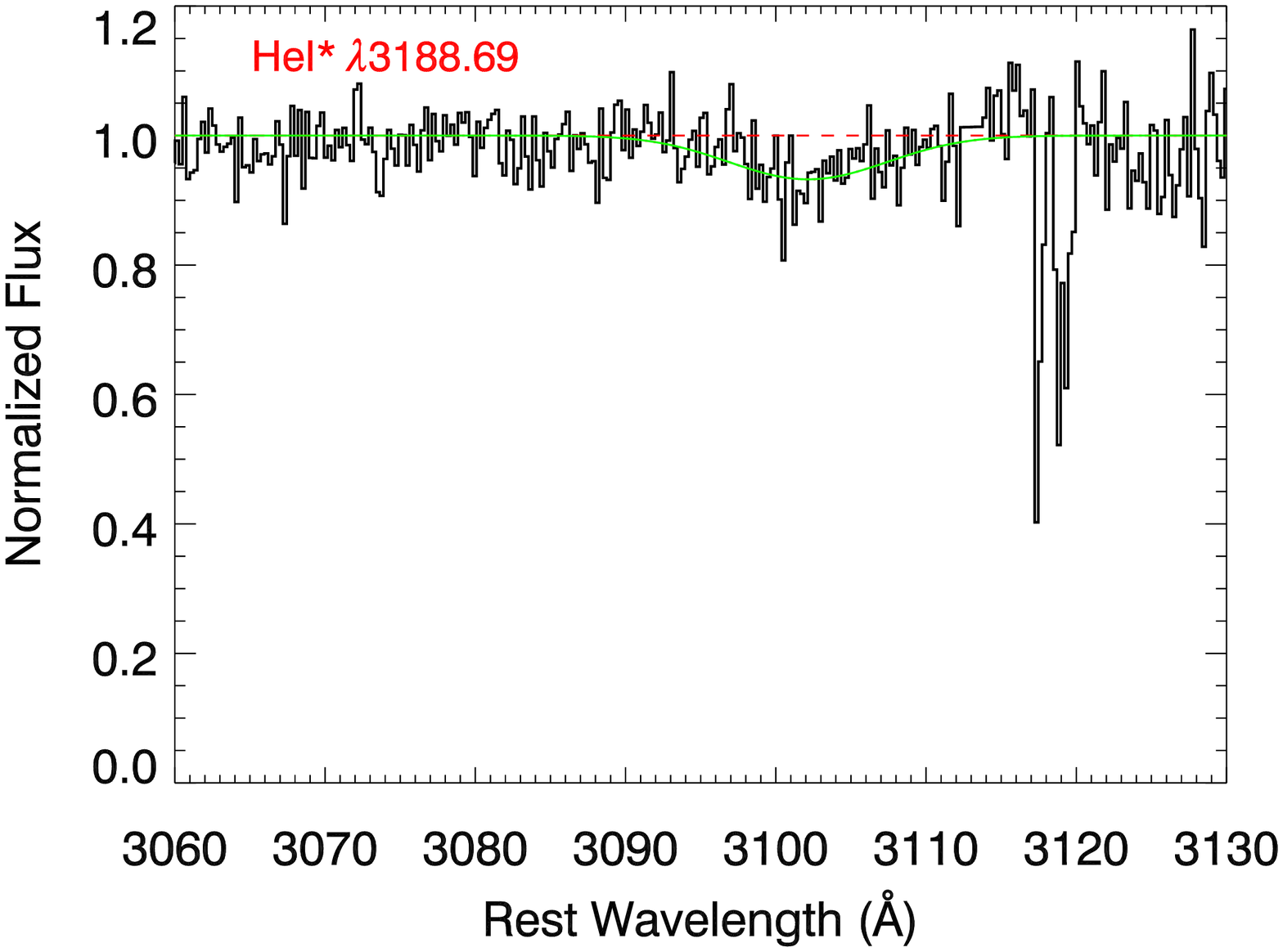}
  \includegraphics[angle=90,width=0.31\textwidth]{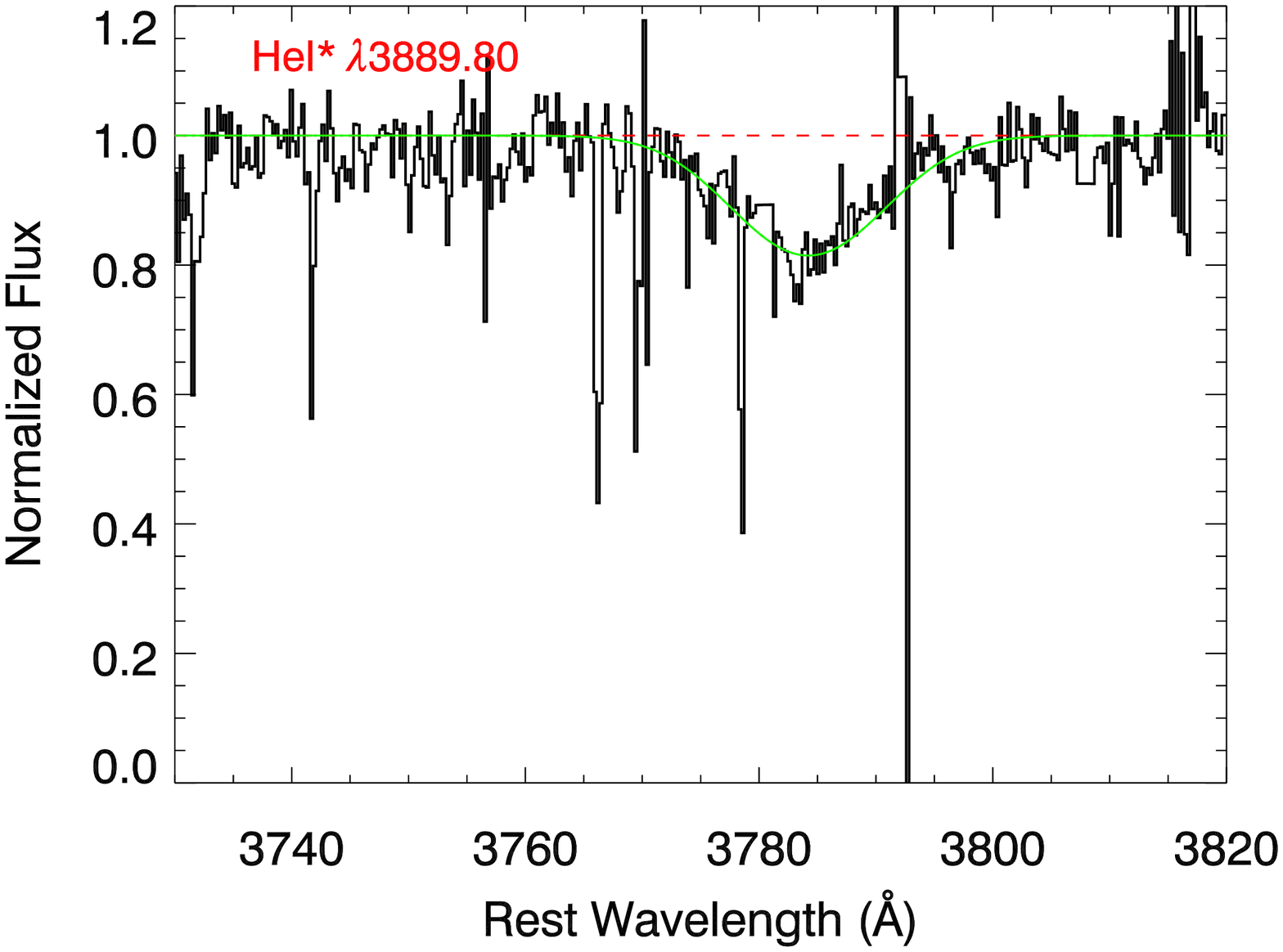}\\
 \caption{Fits to the mildly blended high-ionization absorption troughs (\pv, \siv\ and \siiv), as well as the non saturated troughs
(\hei*, \mgii, \cii, \alii, \aliii) in the X-shooter spectrum of SDSS J1106+1939. Each transition is fitted with an identical 
optical depth profile whose shape is described by a sum of two Gaussians G1+G2 (Model 1, see Section~\ref{identifeat}) leaving only
the maximum optical depths as a free parameter in each fit. G1 is represented by a dotted line and G2 by a dashed line. The full model
for the trough is plotted as a solid line. This simple model provides a good fit to \hei* and the low ionization species as well as
to the unblended blue wing of the \pv\ absorption. The poor fit of the \pv, \siv + \siv* and \siiv\ BAL red wings suggests the existence
of another component that is not seen in the low ionization species (see text and Figure~\ref{figfirstshot2}).}
 \label{figfirstshot} 
\end{figure}

The X-shooter spectrum of SDSS J1106+1939 exhibits wide absorption troughs associated with \hi, \civ, \siiv, \nv,
\ovi, \siv/\siv* and \pv\ ionic species. The \civ\ absorption trough (and most other troughs as well) satisfies
the observational definition of a broad absorption line \citep[BAL, see][]{Weymann91}. In addition, we identify
absorption at the same kinematic location associated with \hei* and the lower ionization species \mgii, \cii,
\alii, and \aliii\ (see Figures~\ref{fullspec1106} and~\ref{figfirstshot}).

The column density of an ionic species $i$ associated with a given kinematic component is estimated by modeling the
residual intensity $I_i(\lambda) \equiv F_{obs}(\lambda)/F_0(\lambda)$ as a function of the radial velocity. The
simplest modeling technique is the apparent optical depth (AOD) method where $\tau_i(v)\equiv -\ln(I_i(v))$,
which is then converted to a column density $N_i(v)$ using the appropriate atomic/physical
constants \citep[see Equation~9 in][]{Savage91}. However, our group \citep{Arav97,Arav99,Arav99ak,Arav01b,Arav01,Arav02,Arav03,Scott04,Gabel05a}
and others \citep{Barlow97a,Hamann97b,Telfer98,Churchill99,Ganguly99} showed that column densities derived from the apparent optical
depth analysis of BAL troughs are unreliable due to non-black saturation in the troughs. In particular, in Paper II we showed that the
true \civ\ optical depth in the deepest outflow component of SDSS J1512+1119, is $\sim$1000 times greater in the core of the
absorption profile than the value deduced from the AOD method.

To account for non-black saturation in unblended doublets or multiplet troughs from the same ion, we routinely use the 
partial covering (PC) and power-law (PL) absorption models \citep[e.g.][for details]{Arav99,Arav08,Edmonds11,Borguet12a}.
However, the intrinsic width of most absorption troughs in the spectrum of  SDSS J1106+1939 causes self-blending of
troughs from most of the observed doublets (\civ, \nv, \ovi, \mgii, \aliii, \siiv\ and \pv). As
a result, the pure PC and PL methods cannot be used. We therefore rely on the template fitting technique that has been
widely used in the study of BAL quasar spectra \citep[e.g.][and references therein]{Korista92,Arav99ak,deKool02b,Moe09}.
The main assumption made when using this technique is that the physical properties of the absorbing gas do not significantly
change as a function of the radial velocity for a given kinematic component \citep[e.g.][]{Moe09,Dunn10a}.
Therefore, the optical depth profile as a function of velocity is assumed to be proportional for all lines so that a single
scaled template is used to reproduce the observed absorption troughs. In the three cases where this assumption was tested 
\citep[][]{Korista08,Moe09,Dunn10a}, it was shown to be well consistent with the data. As we will show here, we obtain good fits to most troughs in
SDSS J1106+1939 using this simple and restrictive assumption. During the fitting procedure, we assume that doublet lines are not affected by
non-black saturation, as a first approximation, and relax that constraint when needed in order to obtain a better
fit to the observed line profile.  It is important to note that in most doublet and multiplet cases, template
fitting can indicate whether the trough is highly saturated or whether the actual column density is close
to the non-saturated AOD case \citep[see][for details]{Moe09}.

We choose the \pv\ $\lambda \lambda$ 1117.98,1128.01 doublet lines as a starting point for the template building.
Our main motivations in the choice of \pv\ are the high velocity separation between its components ($\Delta v \sim$ 2700 km s$^{-1}$), and the absence
of other strong intrinsic absorption and emission lines in that region of the spectrum.
The first step in our template building process consists of matching the blue wing profile of the \pv\ BAL since that part of the trough is
not blended with kinematic components from the red line of the doublet. The absorption structure visible in the \pv\ blue wing suggests that more than
one kinematic component might be present for the blue transition. Therefore, for our first model we choose the optical depth template 
of each line to be a sum of two Gaussians G1 and G2 (Model 1).

In Figure~\ref{figfirstshot} we show the result of the best fit to the moderately blended transitions in the SDSS J1106+1939 spectrum
using Model 1. The FWHM ($w_i$) and central positions ($v_i$) of the two Gaussians have been fit simultaneously to reproduce the
\pv\ BAL troughs. The fit suggests that the blue wing of the \pv\ BAL profile is well
represented by a simple model in which the first Gaussian G1 is characterized by $w_1 = 1150$ km s$^{-1}$, $v_1=-8250$ km s$^{-1}$ and
the second Gaussian G2 by $w_2 = 2400$ km s$^{-1}$, $v_2=-10100$ km s$^{-1}$. An absolute lower limit on the \pv\ column density
present in G1 and G2 is placed by using the AOD technique on the blue transition whose profile is known and unblended. Relaxing the
2:1 constraint of optical depth between the doublet lines of \pv, we are able to obtain a better match of the observed line profile at the location
of the core of G1 \pv $\lambda 1128.01$ (see first panel of Figure~\ref{figfirstshot}) by increasing the optical depth associated with that transition. We place an upper limit
on \pv\ column density in G1 by using the PC technique on this 1:1.1 optical depth ratio doublet (see Table~\ref{coldensmo1}).
This estimation constitutes a hard upper limit since the red \pv\ line of that component is likely blended by the \pv\ blue transition
of a lower velocity system (see Model 2) such that the true ratio of optical depth for that system is within 1:2 to 1:1.1.

The optical depth model derived from the \pv\ BAL fit is also able to reproduce a significant part of the blended \siv/\siv* BAL.
The \siv/\siv* model presented in Figure~\ref{figfirstshot} considers the presence of the resonance $\lambda 1062.97$ and excited
$\lambda 1073.01$ line (the oscillator strength weighted mean of the blend of the two close-by $E_{low} = 951$ cm$^{-1}$ \siv\
transitions, see paper II for details) for which $\Delta v \sim$ 2900 km s$^{-1}$. Due to the uncertainty in the \ovi\ emission
contribution in the blue wing of \siv\ (see Section~\ref{unabmod}), the column density associated with component G2 for that transition is less certain
and could be underestimated by up to a factor of two if we would not have scaled down the shifted \civ\ emission model. 
Model 1 is also able to reproduce the absorption profile observed in the other species such as \hei*, \mgii, \cii, \alii, and
\aliii\ as can be seen in the remaining panels of Figure~\ref{figfirstshot} in which the maximum optical depths are the only
free parameters of the models. The fit of these profiles was performed without having to relax the 2:1 assumption for the ratio of optical
depth between the strongest and weakest doublet components, consistent with non saturated troughs. We report the column densities derived for each ionic species in Table~\ref{coldensmo1}. In \hei*, we do not detect any
absorption related to component G2 and report an upper limit on the column density for that component by scaling the Gaussian template
assuming that that the noise level could hide a $2 \sigma$ detection for the strongest \hei* $\lambda 3889.80$ transition.
The \civ, \siiv, \ovi, \nv\ doublet are both more heavily self-blended than \pv\ and \siv\ because of the smaller velocity
separation between the doublet components, and are much more saturated due to their higher elemental abundance. This prevents
us from obtaining reliable fits to their absorption troughs, and only lower limits on their column densities can be derived.
Blending of the \nv, \hi\ and  \siIII\ troughs prevents us from deriving useful column densities for these species as well (beside
the fact that \nv, \hi\ must be heavily saturated). We however
note that the lower limits derived from the AOD technique for these species are consistent with the predicted values in our
photoionization models. We place an upper limit on the column density of the non detected \feii\ by scaling the template of each Gaussian
to within $2\sigma$ at the expected location of the \feii\ $\lambda$1608.45 troughs. We did not use the stronger \feii\ $\lambda$2600.17
that is located in a region of poor SN at the blue edge of the NIR detector or the \feii\ $\lambda$2382.76 that is severely blended
by a H$_2$O atmospheric band.

\begin{deluxetable}{lccccc}
 \tablewidth{0.95\textwidth}
 \tablecolumns{6}
 \tabletypesize{\footnotesize}
 \tablecaption{Ionic column densities in the outflow of SDSS J1106+1939 using Model 1}
 \tablehead{
 \colhead{Ion} &
 \multicolumn{2}{c}{G1 (-8250 km s$^{-1}$)} &
 \multicolumn{2}{c}{G2 (-10100 km s$^{-1}$)} &
 \colhead{Adopted$^\mathrm{a}$}\\
 \colhead{} &
 \colhead{AOD} &
 \colhead{PC} &
 \colhead{AOD} &
 \colhead{PC} &
 \colhead{(G1+G2)}\\
 \colhead{} &
 \colhead{($10^{12}$ cm$^{-2}$)} &
 \colhead{($10^{12}$ cm$^{-2}$)} &
 \colhead{($10^{12}$ cm$^{-2}$)} &
 \colhead{($10^{12}$ cm$^{-2}$)} &
  \colhead{($10^{12}$ cm$^{-2}$)} 

  }
 \startdata

 \hei*  &   450$^{+40}_{-30}$    &  480$^{+200}_{-20}$   &   $<140$            &  ...                      &  480$^{+340}_{-20}$     \\
 \cii   &   510$^{+30}_{-20}$    &     ...            &   220$^{+40}_{-20}$      & ...                     &  730$^{+50}_{-30}$       \\
 \cii*   &  980$^{+30}_{-20}$    &    ...                & 530$^{+40}_{-30}$     & ...                     & 1500$^{+50}_{-40}$       \\
 \alii  &  16$^{+1}_{-1}$    &   ...                & $< 7.1$                & ...                   & 16$^{+8}_{-1}$     \\
 \aliii &  240$^{+10}_{-10}$        &  250$^{+10}_{-10}$   & 180$^{+10}_{-10}$     & 180$^{+15}_{-10}$           &  430$^{+20}_{-10}$       \\
 \pv\   &  1300$^{+30}_{-20}$   &   4500$^{+900}_{-150}$  & 1900$^{+30}_{-30}$   & 2700$^{+1700}_{-50}$    & 3200--9900$^\mathrm{b}$  \\
 \siv\  & 38000$^{+22000}_{-1100}$ $^\mathrm{c}$    &  ...   &  5200$^{+100}_{-100}$         & ...         & 44000$^{+22000}_{-1100}$ \\
 \siv*  & 16000$^{+300}_{-200}$     &   ...               &   2500$^{+100}_{-100}$      & ...               & 19000$^{+300}_{-200}$    \\
 \mgii   &   140$^{+10}_{-10}$    &  140$^{+10}_{-10}$      &      84$^{+5}_{-5}$   &  85$^{+36}_{-3}$        & 290$^{+40}_{-10}$         \\
 \feii   &  $< 160$              & ...                  & $< 370$                & ...                     & $< 540$                  \\

 \enddata


 \label{coldensmo1}   
 a) Column density values adopted for photoionization modeling (see Section~\ref{photocon}).\\
 b) The range translates the uncertainty on the \pv\ column density due to the possible non black saturation affecting component G1 (see text).\\
 c) The upper error for the black \siv\ G1 component has been computed by finding the maximum optical depth that can be
hidden until the wings of component G1 become too broad for the observed line profile.
\end{deluxetable}

\begin{figure}
  \includegraphics[angle=90,width=0.5\textwidth]{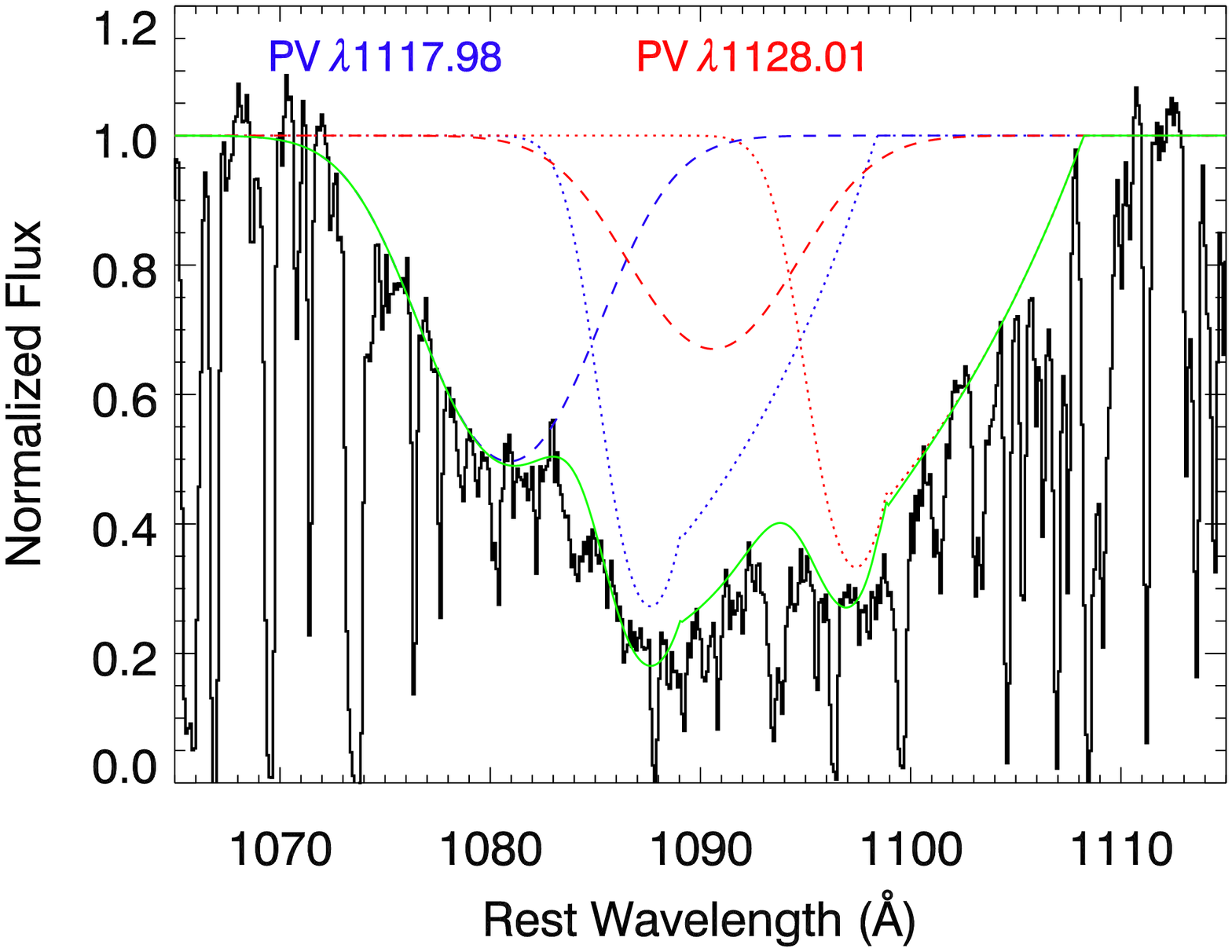}
  \includegraphics[angle=90,width=0.5\textwidth]{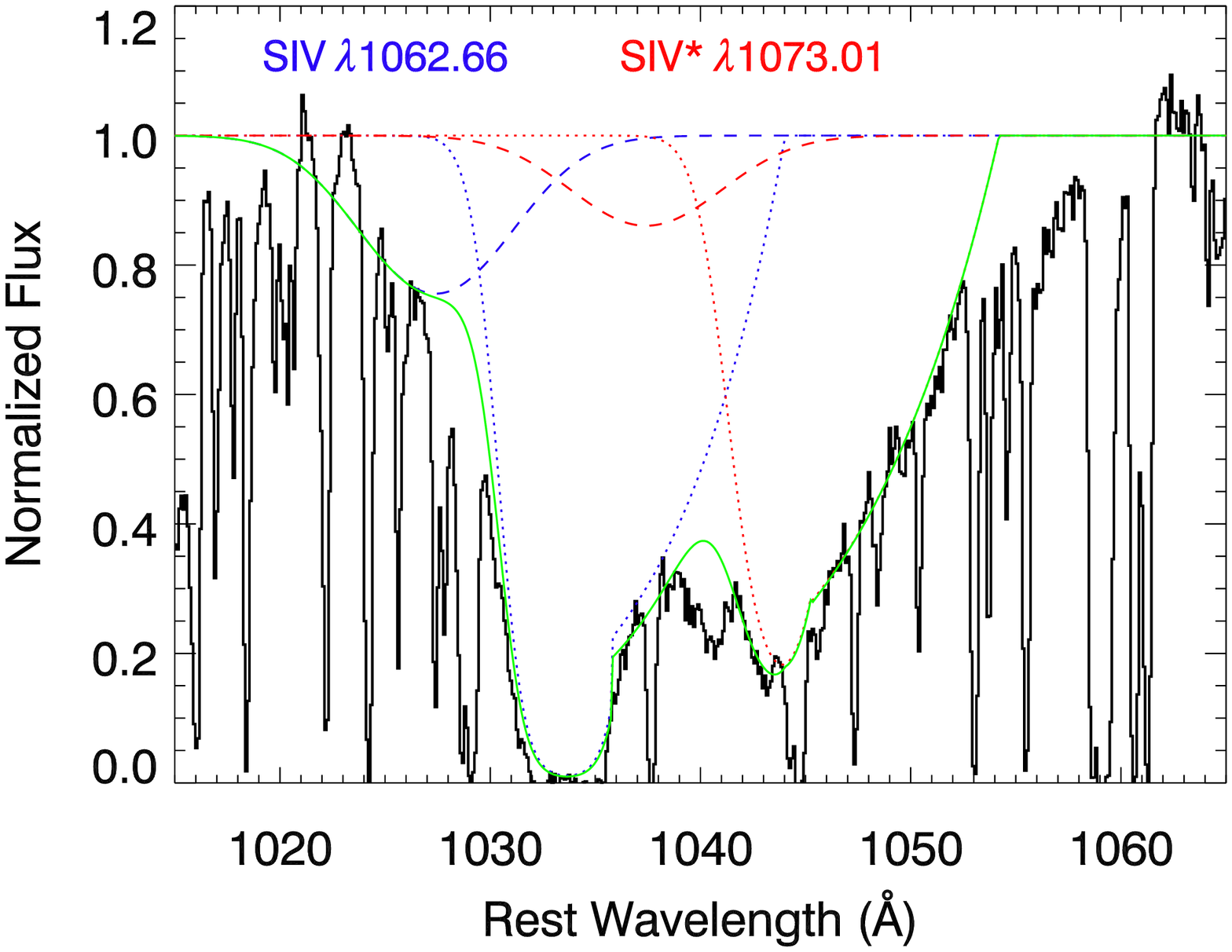}\\
  \includegraphics[angle=90,width=0.5\textwidth]{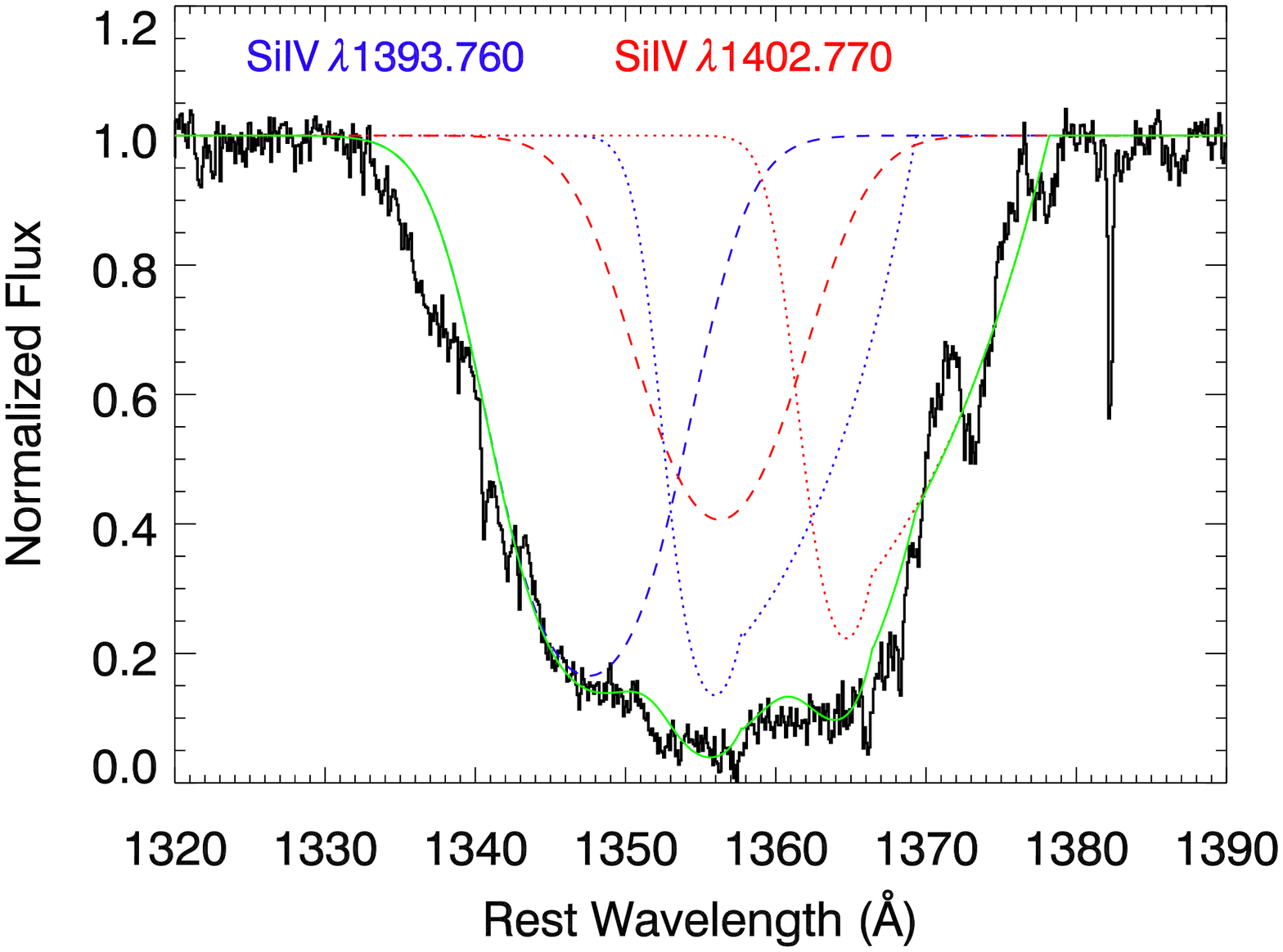}
 \caption{Fitting the high ionization \pv, \siv +\siv* and \siiv\ BAL profiles with the Model 2 optical depth profile. As for Model 1,
the optical depth distribution associated with a transition is assumed to be known, leaving only
the maximum optical depths of each components as a free parameter in each fit. Model 2 is composed of a sum of two functions: F1
(a Gaussian with a modified red wing) represented by a dotted line and F2 (a Gaussian with identical parameters as G2) represented
by a dashed line. The total model for the trough is plotted as a solid line. For \pv\ the fit requires a ratio of optical depth
between both components of the doublets of 1:1.1, suggesting the possibility of moderate non-black saturation.}
 \label{figfirstshot2} 
\end{figure}

While Model 1 provides a reasonable fit to the observed absorption profile associated with most of the
ionic species, it fails to reproduce the red wing observed in the BAL profile of high ionization species
such as \pv, \siv + \siv* or \siiv\ (see Figure~\ref{figfirstshot}). This observation suggests the existence
of an additional kinematic component associated with the high ionization species. In order to better fit the
red wing of the high ionization lines, we built a second template, Model 2. This second optical depth template is also composed of two functions: a high velocity
Gaussian F2 with identical parameters as G2, and F1: a low velocity modified Gaussian profile. The result
of the simultaneous fit, performed with the same constraints as with Model 1, is presented for \pv, \siv + \siv* and \siiv\ in
Figure~\ref{figfirstshot2}, and the associated column densities are shown in Table~\ref{coldensmo2}.
Here the overall fit of the \pv\ BAL is much better, as well as the fit to the \siv +\siv* profile.
The reported column densities are consistent within $\sim$30\% with the one derived using Model 1.
While Model 2 provides a better fit to the high ionization \pv\ and \siv + \siv* BAL (but also \siiv),
the lack of constraints on the shape of the low velocity component template makes the derived column
densities too model dependent to be reliable. Moreover, the small fractional difference in column density from that component does not significantly
affect the photoionization modeling since the bulk of the computed column density is located in the two main
components G1 and G2. For these reasons, we will use the column densities reported in Table~\ref{coldensmo1}.

\begin{deluxetable}{lcccc}
 \tablewidth{0.95\textwidth}
 \tablecolumns{5}
 \tabletypesize{\footnotesize}
 \tablecaption{Ionic column densities in the outflow of SDSS J1106+1939 using Model 2}
 \tablehead{
 \colhead{Ion} &
 \multicolumn{2}{c}{F1 (-8250 km s$^{-1}$)} &
 \multicolumn{2}{c}{F2 (-10100 km s$^{-1}$)}\\
 \colhead{} &
 \colhead{AOD} &

 \colhead{PC} &
 \colhead{AOD} &
 \colhead{PC} \\
 \colhead{} &
 \colhead{($10^{12}$ cm$^{-2}$)} &

 \colhead{($10^{12}$ cm$^{-2}$)} &
 \colhead{($10^{12}$ cm$^{-2}$)} &
 \colhead{($10^{12}$ cm$^{-2}$)} 

  }
 \startdata

 \pv\   &  3300$^{+40}_{-30}$     & 5700$^{+3200}_{-110}$ & 1800$^{+30}_{-30}$      & 2400$^{+320}_{-30}$        \\
 \siv\  & 47000$^{+4200}_{-1200}$  &  ...                 &  5500$^{+560}_{-140}$         & ...        \\
 \siv*  & 26000$^{+300}_{-300}$     &   ...               &   2900$^{+530}_{-130}$      & ...        \\

 \enddata


 \label{coldensmo2}
\end{deluxetable}

We presented two phenomenological decomposition of the BAL profiles that allows us to estimate the column densities associated with \siv\ and \siv*
to better than within 50\% (see Table~\ref{coldensmo1} and~ \ref{coldensmo2}). The larger uncertainty affecting the \siv\ column density in the
resonance state has been estimated by scaling the G1 template so that the wings of the model are no longer consistent with the observed profile. 
In the last column of Table~\ref{coldensmo1}, we report the adopted column densities for the photoionization modeling, as well as the statistical error affecting the measurements (taking only the photon noise into account). For photoionization modeling we choose to sum components G1 and G2 for the following reason:
The ratio of almost all ionic column density of the same ions is roughly 2:1 between G1 and G2, suggesting very similar photoionization
solution. The exception is \siv\ where the ratio is $\sim7$. However as we noted above, this is mainly due to our choice of a minimal \ovi\
BEL. If we choose an \ovi\ BEL as strong as the \civ\ BEL the column density of \siv\ doubles and triples if we choose a 1.5 times stronger \ovi\ BEL. 
This will make the \siv\ ratio between components G1 and G2 $\sim$3.5 and $\sim2$ respectively.  We therefore conclude that it is highly
plausible that both components arise from the same outflow and have similar ionization equilibria, which justify adding them together
for the purpose of photoionization modelling and the extraction of the physical parameters of the outflow. When available, we choose to use the value reported
in the PC column as the measurement. If only AOD determination is available, we will consider the reported value minus the error as a lower
limit during the photoionization analysis since conceptually no information about non-black saturation effects can be obtained from singlet lines.
For \pv, due to uncertainty in the column density associated with component G1, we report the range of values that are allowed within the absolute
lower and upper limits placed on the column density.

The template decomposition also demonstrates the detection of a \siv* BAL 
outflow and the fact that the column density associated with the excited state is lower than the one measured in the resonance level. 
The detection of the \siv* BAL is secured by the good kinematic match of the template fit.
Furthermore, if that part of the profile would have been related to another low velocity \siv\ system rather than \siv*, then no kinematic
match would be observed with not only \pv, but also with higher abundance species like \civ, \ovi\ or \nv\ that would have likely
been heavily saturated.

We note that the observation of absorption associated with the low ionization \aliii\ and \siIII\ lines
suggests the presence of absorption due to the \feiii $\lambda$ 1122.524 blending with the \pv\ BAL line. Such a blend
could affect the \pv\ column density derived during our template fitting procedure if the \feiii\ column density was significant
enough. The photoionization model presented for the SDSS J1106+1939 outflow in Section~\ref{photocon} predict an
optical depth $\tau < 0.1$ for that \feiii\ line, which does not affect our \pv\ column density determination.

\begin{deluxetable}{lcc}
 \tablewidth{0.7\textwidth}
 \tablecolumns{3}
 \tabletypesize{\footnotesize}
 \tablecaption{Ionic column densities associated with component 4 of the SDSS J1512+1119 outflow.}
 \tablehead{
 \colhead{Ion} &
 \colhead{AOD} &
 \colhead{PC}\\
 \colhead{} &
 \colhead{($10^{12}$ cm$^{-2}$)} &
 \colhead{($10^{12}$ cm$^{-2}$)} 

  }
 \startdata

   \hi   &   $\geq 1427^{\mathrm{a}}$    & ...                \\
  \nv   &   $\geq 1270$                 & ...                \\
  \siIII &  $\geq 5.7 $                 & ...                \\
  \siiv  &  $72.1^{+1.0}_{-1.0}$        &  $119^{+18}_{-6}$  \\
  \siv   &  $\geq 263 $                 &                    \\
  \siv*   & $\leq 25  $                    & ...                \\

 \enddata


 \label{tabj1512_t4}

 a) Obtained through AOD modeling of the Ly$\beta$ absorption profile.
\end{deluxetable}

\subsection{Column Density in SDSS J1512+1119}

\begin{figure}
  \includegraphics[angle=90,width=1.0\textwidth]{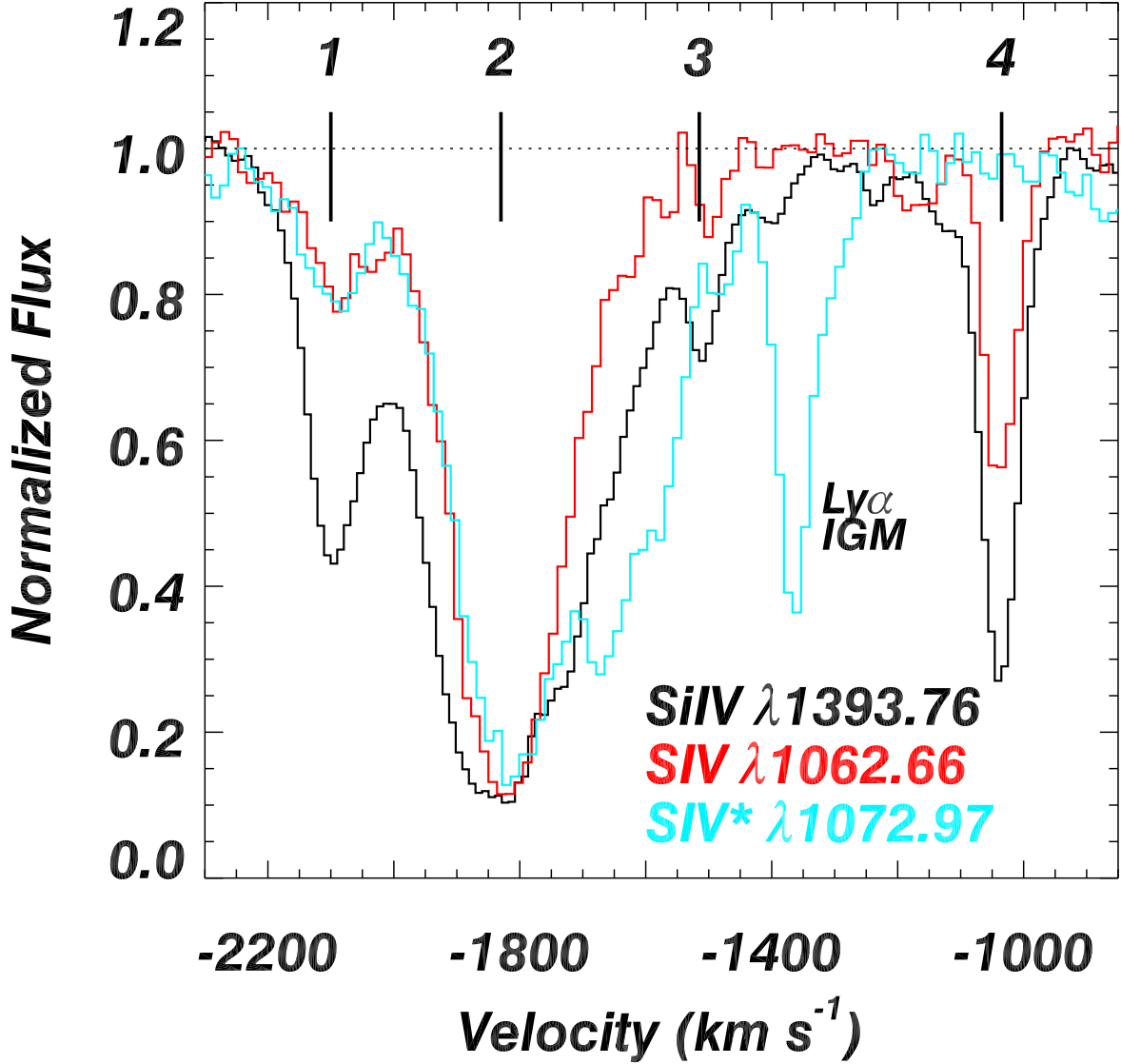}\\
 \caption{Identification of the \siv\ and \siv* intrinsic absorption troughs in the spectrum of SDSS J1512+1119. The red
wing of the troughs associated with component 2 of \siv* $\lambda 1072.97$ is mainly affected by a blend due to the \siv* $\lambda 1073.51$
transition of component 2, that allowed us to pinpoint the \siv* column density for that component (see Paper II).}
 \label{figj1512siv} 
\end{figure}

Our second target, SDSS J1512+1119, displays a \civ\ absorption trough whose width is close to a BAL
($\Delta v \sim 1700$ km s$^{-1}$, see Paper II). However, the \siiv\ lines are narrow enough and unblended ($\Delta v \sim 2000$ km s$^{-1}$) to be
adopted as a template to identify the kinematic components in other ionic species. Using that template,
we reported in Paper II the existence of five main kinematic components with centroids at velocities
of $v \sim$ -2100, -1850, -1500, -1050, and -520 km s$^{-1}$ (labeled component 1--5). In this paper
we concentrate on determining the distance and energetics associated with the two main components:
component 2 and 4.
As shown in Figure~\ref{figj1512siv} in which we use the \siiv\ $\lambda 1393.76$ line as template, absorption
troughs associated with \siv\ and \siv* are observed in kinematic component 2 ($v \sim$ -1850 $^{-1}$) and \siv\
absorption is also detected in  kinematic component 4 ($v \sim$ -1050 $^{-1}$). The in depth analysis of the \siv\ and \siv*
troughs performed in Paper II revealed that the blend affecting the red wing of the component 2 in \siv* $\lambda 1072.97$ is
identified as the ten times weaker transition of \siv* $\lambda 1073.51$. This allowed us to derive reliable estimates
of $N($\siv) $\sim 10600^{+4000}_{-2700} \times 10^{12}$ cm$^{-2}$ and $N($\siv*) $\sim 16000^{+5000}_{-1300} \times 10^{12}$ cm$^{-2}$
for that component.

For component 4, we estimate the \siv\ column density by using the AOD model on the
normalized line profile (i.e. $\tau(v) = -\ln (I_i(v))$), converting it to column densities and integrate it over the trough
to obtain a lower limit of $N($\siv) $\geq 263 \times 10^{12}$ cm$^{-2}$. We do not detect an absorption trough associated with \siv* for
that component, so we report an upper limit on the column density for that transition by scaling the \siv\ template, assuming
that the noise could hide up to a 2 $\sigma$ detection and find $N($\siv*) $\leq 25 \times 10^{12}$ cm$^{-2}$.
Due to the narrowness of the line profile and absence of significant blending, we are able to measure the \siiv\ column density
and place limits on the ionic column for other species that we report in Table~\ref{tabj1512_t4}. The constraints derived from
the \civ\ and \ovi\ column densities are consistent with the one obtained from the \nv\ and are not reported in the table. We finally
place an upper limit on the \pv\ column density of $N$(\pv)$< 8.1 \times 10^{12}$ cm$^{-2}$ due to its non detection in that component. 


\section{PHOTOIONIZATION ANALYSIS}
\label{photocon}

We use photoionization models in order to determine the ionization equilibrium of the outflow, its
total hydrogen column density ($N_H$), and to constrain its metallicity. The ionization parameter
\begin{equation}
\label{equauh}
 U_H\equiv\frac{Q_H}{4 \pi R^2 c \vy{n}{H}},
\end{equation}
(where $Q_H$ is the source emission rate of hydrogen ionizing photons, $R$ is the distance to the absorber from
the source, $c$ is the speed of light, and $\vy{n}{H}$ is the hydrogen number density) and $N_H$ of the
outflow are determined by self-consistently solving the ionization and thermal balance equations with
version c08.00 of the spectral synthesis code Cloudy, last described in \citet{Ferland98}. We assume a
plane-parallel geometry for a gas of constant $\vy{n}{H}$ and initially choose solar abundances as given
in \citet{Lodders09}. Given the lack of observational constraints in wavebands outside the X-shooter spectral
range for both objects, we choose the UV-soft spectral energy distribution (SED) model for high luminosity
radio quiet quasars described in \citet{Dunn10a}. The use of this SED model, lacking the so-called big
blue bump from the classical MF87 SED \citep{Mathews87}, is motivated by the rather soft FUV slopes observed
by \citet{Telfer02} over a large sample of HST spectra of typical radio quiet quasars \citep[see][for a
detailed discussion]{Dunn10a}. Using this SED, we generate a grid of models by varying $N_H$ and $U_H$.
Ionic column densities predicted by the models are tabulated and compared with the measured values in order
to determine the models that best reproduce the data.

\subsection{SDSS J1106+1939}
\label{pij1106}

We compare the ionic column densities given in the last column of Table~\ref{coldensmo1}
(see discussion in Section~\ref{identifeat}) with predictions of photoionization models, and use
$\vy{n}{H}=10^{4}$~cm$^{-3}$, which we measure for the outflow using the ratio of \siv/\siv* ionic
column densities (see Section~\ref{densideter}). We focus our modeling on the troughs from the high
ionization species \hei*, \pv, and \siv, as these are the dominant species in the outflow, and \siv\ is
the species that allows us to constrain $R$ (see Section~\ref{densideter}). We treat
\hei* as a high ionization species since its concentration depends linearly on the fraction of \heii\
in the gas \citep[see e.g.][]{Arav01b}.

\begin{figure}
  \includegraphics[angle=00,width=0.95\textwidth]{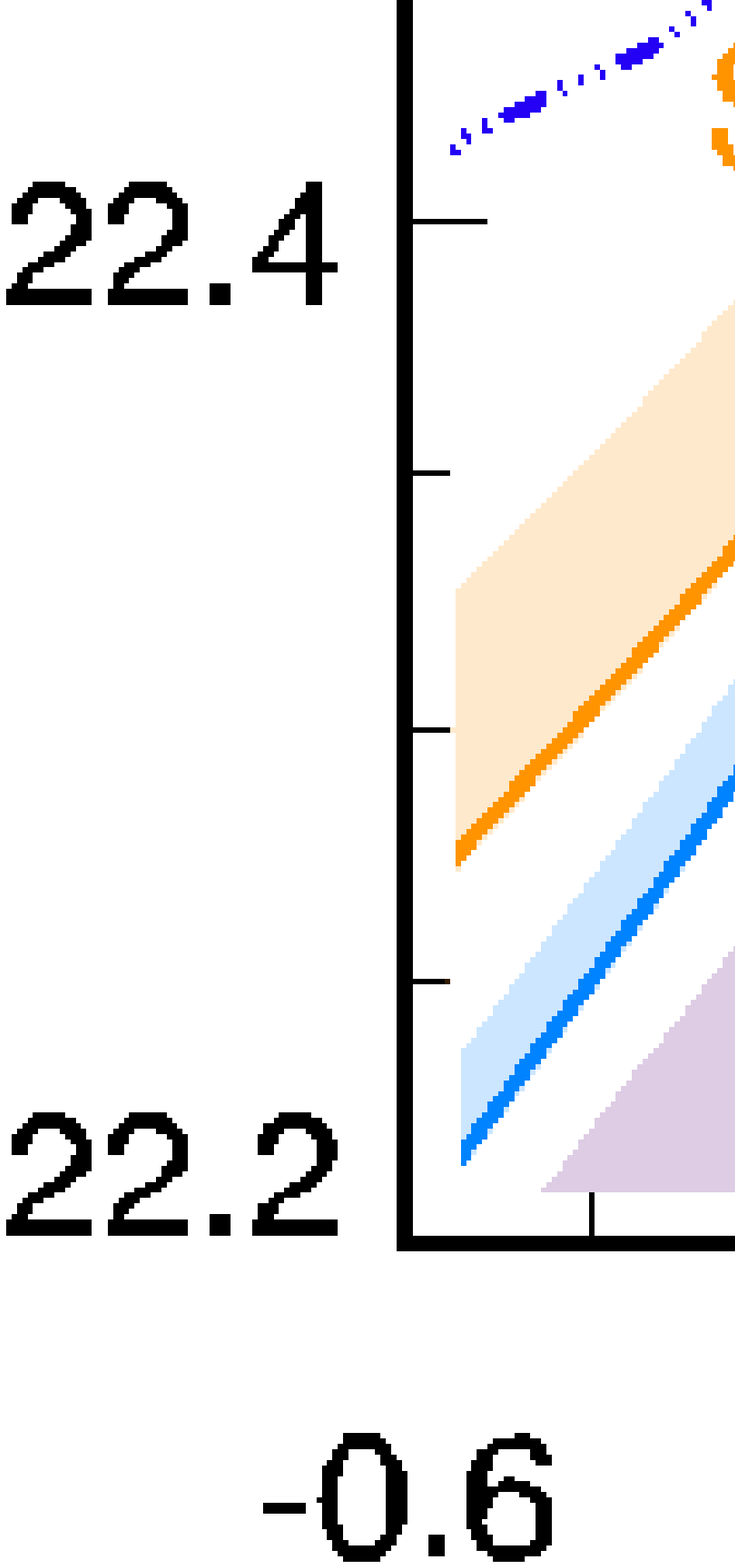}
 \caption{Photoionization modeling of the outflow in SDSS J1106+1939 assuming solar abundances. The shaded
regions represents the locus of points ($U_H,N_H$) able to reproduce the observed ionic column density for a given species (see Table~\ref{coldensmo1}); dashed lines represent an upper limit on the
column density of the ion; and dotted dashed lines represent a lower limit on the column density. The best model is marked by a diamond and the systemic
error on the solution is represented by the cross. For clarity we do not represent the \alii\ line
since it is co-spatial with the \cii\ line.}
 \label{pisol1} 
\end{figure}

In Figure~\ref{pisol1}, contours where model predictions match the measured ionic column densities are plotted
in the $N_H-U_H$ plane. Note that we use the average of the range of
values given for \pv\ in component G1 and include the
upper and lower limits in the error to reflect the uncertainty on the
exact column density in that component
(see Section~\ref{identifeat}). The best-fit model is parametrized by $\log(U_H) = 0.0 \pm 0.2$, and
$\log(N_H)=22.8 \pm 0.2$~cm$^{-2}$. The errors are determined
by the region where \pv\ and \siv\ bands cross and are correlated,
with higher ionization parameters corresponding to higher
total column densities. Comparison of measured and predicted ionic
column densities are given in Table~\ref{tab:modeled}.
As can be seen from Table~\ref{tab:modeled}, the best model predicts
the column densities of the high ionization species to within a factor
of 2. The poor fit to \cii\ and \aliii\ is not physically troubling for the
following reasons: First, as stated above, we purposefully attempted to find the best fit for
the dominant high ionization species in the outflows. Second, within
the reported error bars for $N_H-U_H$ we can find solutions that yield a much better
fit for these two ions. For example, a model with $\log(U_H) = -0.2$ and
$\log(N_H)=22.85 $~cm$^{-2}$ produces both the \cii\ and \aliii\ column
densities to better than a factor of two, with only a moderate worsening
of the high ionization lines fit. The drastic changes in the column
densities of the singly ionized species is due to the proximity of the
solution to the hydrogen ionization front \citep[see][]{Korista08}.

\begin{figure}
  \includegraphics[angle=00,width=0.95\textwidth]{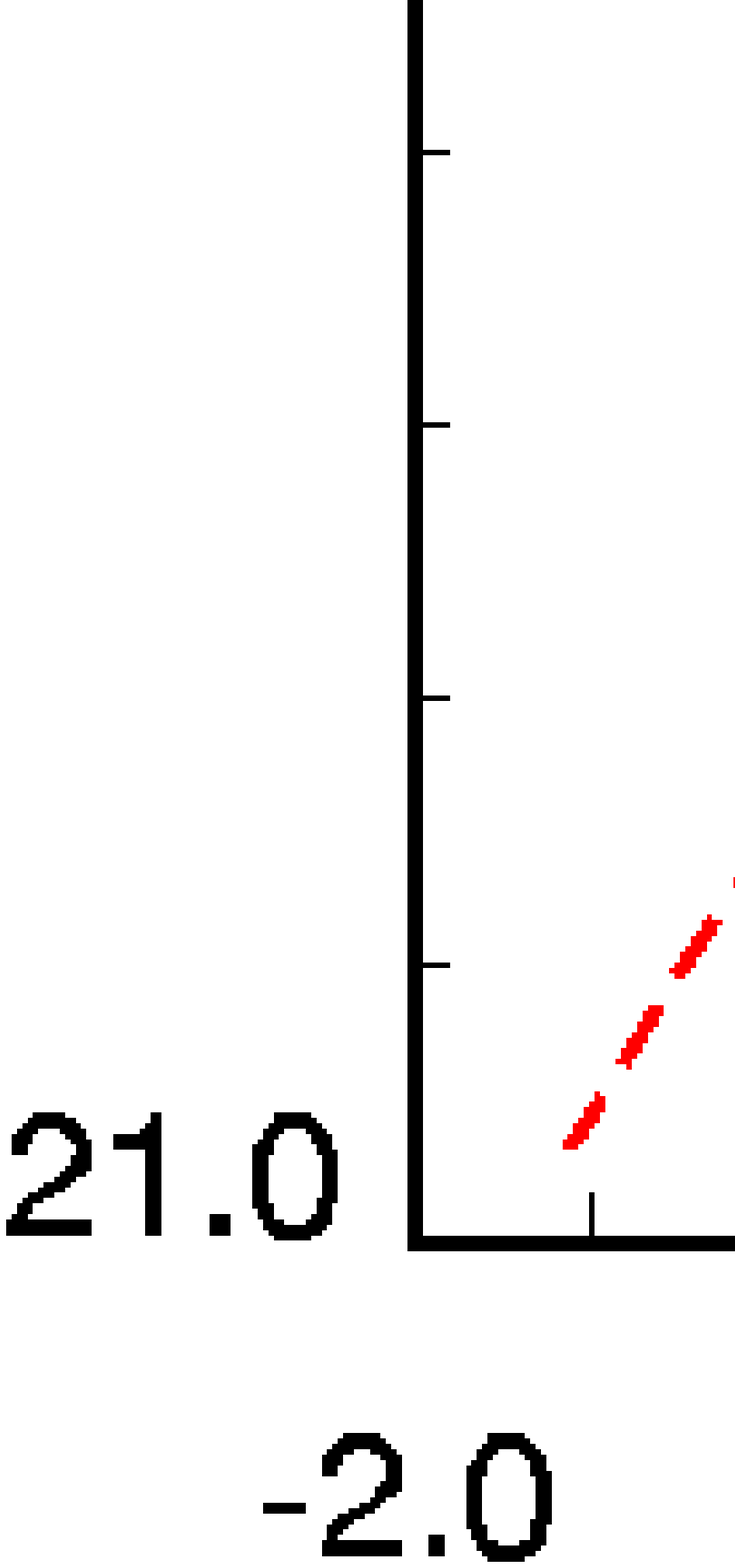}\\
 \caption{Photoionization modeling of the outflow in SDSS J1106+1939 assuming a metallicity $Z/Z_{\odot}=4$.
The description of the figure is identical to Figure~\ref{pisol1}.
   }
 \label{piz41} 
\end{figure}

The analysis above relies on the assumption of solar abundances. However, AGN outflows are known to have
moderate supersolar metallicities (e.g., QSO J2233-606: $Z \simeq 4 Z_\odot$, \citealt{Gabel06}; Mrk 279:
$Z \simeq 2 Z_\odot$, \citealt{Arav07}; SDSS J1512+1119 $Z_\odot \ltorder Z \ltorder 4 Z_\odot$, Paper II).
We therefore investigate how sensitive our results are to higher metallicity models. Comparing the \siv\ line
on the grid models to that of \hei*, we find that the abundance of sulphur must be $\ltsim 4$ times its solar
value. This is because the abundance of helium is relatively insensitive to changes in the metallicity of the
plasma (within $\approx 8-13$\% of hydrogen for $Z \ltsim 4 Z_\odot$), and \hei* is underpredicted compared
with \siv\ for higher sulphur abundances. We therefore run a grid of models using a metallicity of 4 times
solar, which is consistent with the results of the works cited above for other AGN outflows. Our elemental
abundances for the supersolar metallicity model are determined by scaling C, N, O, Mg, Si, Ca, and Fe as
in \citet{Ballero08} while the remaining metals are scaled as in Cloudy starburst models. The computed
models are shown in Figure~\ref{piz41} in which the best-fit model for $Z = 4 Z_\odot$ is characterized by
$\log(U_H) = -0.5^{+0.3}_{-0.2}$, and $\log(N_H) =22.1^{+0.3}_{-0.1}$~cm$^{-2}$. The quoted errors on $U_H$
and $N_H$ are determined in a similar fashion to the solar metallicity case. We consider this model to be the most physically plausible
for SDSS J1106+1939, while at the same time it provides a more conservative (lower) estimate for the kinetic 
luminosity of this outflow (see Section~\ref{firstorder}). In Section~\ref{sensitivity}, we explore the sensitivity of
the photoionization solution to a different SED.

\begin{deluxetable}{lcc}
 \tablewidth{0.95\textwidth}
 \tablecolumns{3}
 \tabletypesize{\footnotesize}
 \tablecaption{Results of the photoionization modeling for SDSS J1106+1939}
 \tablehead{
 \colhead{Model} &
 \colhead{$Z_\odot$} &
 \colhead{$4~Z_\odot$}\\
 \hline
 \colhead{Ion} &
 \colhead{$\log \big(\frac{N_{mod}}{N_{obs}} \big)$} &
 \colhead{$\log \big(\frac{N_{mod}}{N_{obs}} \big)$} 

  }
 \startdata

\hei*  &  0.37 &  0.08 \\
\cii   & -1.83 & -1.32 \\ 
\alii  & -1.81 & -0.74 \\
\aliii & -0.58 & -0.02 \\
\mgii  & -0.09 &  0.11 \\
\pv    & -0.04 & -0.09 \\
\siv   & -0.03 &  0.34 \\

 \enddata


 \label{tab:modeled}

\end{deluxetable}

\subsection{SDSS J1512+1119}

Component 2 of the outflow of SDSS J1512+1119 was analyzed in Paper II. We found $\log(U_H) \approx -0.9^{+0.1}_{-0.1}$ and 
$\log(N_H) \approx 21.9^{+0.1}_{-0.1}$ for the UV-soft SED. Using the SED developed in Mathews \& Ferland (1987), we found the ionization 
parameter and column density dropped by $\approx 0.2$ and $0.3$~dex, respectively. We determined the metallicity of the gas
was approximately solar with an upper limit of about 4 times solar.

\begin{figure}
  \includegraphics[angle=00,width=0.95\textwidth]{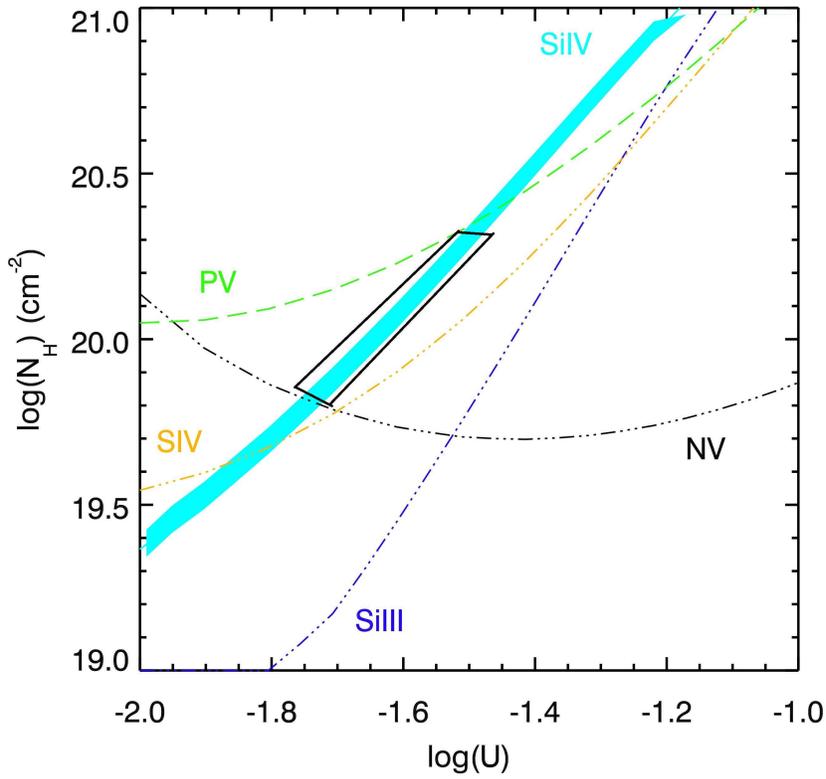}
 \caption{Photoionization modeling of component 4 of the outflow in SDSS J1512+1119 assuming solar abundances. As in Figure~\ref{pisol1}, a measured
column density is represented by a shaded area (showing the uncertainties), an upper limit on a column density is represented by a dashed line and a lower limit
by a dotted dashed line. The region of the phase space in which models are able to reproduce the observed constraints is roughly delimited by the trapezium.}
 \label{pisol2} 
\end{figure}

For component 4 of this outflow, \siiv\ is the only ion for which we have a column density measurement. However, as seen 
in Figure~\ref{pisol2}, the upper and lower limits of other ions help constrain the models. Lower limits on \nv\ and 
\siv\ along with the upper limit on \pv\ constrain the solution to be located within the region interior to these lines in Figure~\ref{pisol2}. 
The solutions then lie along the the \siiv\ band within that region, with $-1.7 \ltsim \log(U_H) \ltsim -1.4$ and 
$19.8 \ltsim \log(N_H) \ltsim 20.4 ~(\rm{cm}^{-2})$.

\section{ESTIMATING THE DISTANCE AND ENERGETICS OF THE \siv\ OUTFLOWS}
\label{firstorder}

Assuming that the absorbing material can be described as a thin ($\Delta R /R \ll 1$), partially-filled shell,
the mass flow rate ($\dot{M}$) and kinetic luminosity ($\dot{E}_k$) of the outflow are given by (see discussion
in \citealt{Borguet12a}):
\begin{eqnarray}
  \dot{M} &=& 4 \pi R \Omega \mu m_p N_{H} v \\
  \dot{E_k} &=& 2 \pi R \Omega \mu m_p N_{H} v^3,
  \label{ekeq}
\end{eqnarray}
where $R$ is the distance of the outflow from the central source, $\Omega$ is the global covering fraction of the outflow,
$\mu$ = 1.4 is the mean atomic mass per proton, $m_p$ is the mass of the proton, $N_H$ is the total hydrogen column density
of the absorber, and $v$ is the radial velocity of the kinematic component, which is directly derived from the trough's profile.
In the rest of this section, we detail the determinations of (or constraints on) $R$ and $\Omega$ that are needed for calculating
$\dot{M}$ and $\dot{E}_k$.

\subsection{Determining $R$: the Distance of the Outflow from the
Central Source}
\label{deterer}

Measuring $R$ is crucial for estimating of $\dot{M}$ and $\dot{E}_k$, and is also essential for understanding the
relationship of the outflows with the host galaxy and its surroundings. In Section~\ref{photocon}, we derived the
ionization parameter ($U_H$) for each outflow. Therefore, a knowledge of $\vy{n}{H}$ and $Q_H$ allows us to solve for
$R$ directly from Equation~(\ref{equauh}).

\subsubsection{Determining $\vy{n}{H}$}
\label{densideter}

\begin{figure}
  \includegraphics[angle=90,width=1.0\textwidth]{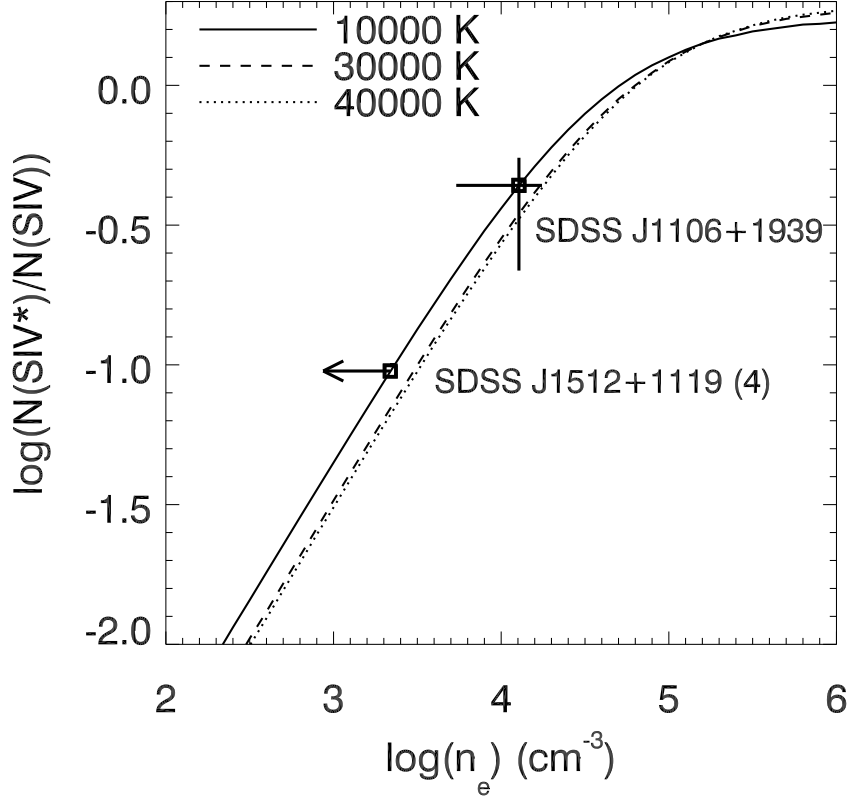}\\
 \caption{Density diagnostic (and errors) for SDSS J1106+1939 and kinematic
component $4$ of the SDSS J1512+1119 outflow. We plot the theoretical
ratio of the level population
of the first excited states of \siv\ ($E=951~ \mathrm{cm}^{-1}$) to
the level population of the
ground state versus the electron number density $n_e$ for three
representative temperatures. The
density diagnostic provided is fairly temperature independent for the
range of densities represented
in this plot. We determine the electron number density associated with
the two outflows using the
curve for the temperature of 10,000 K found for our best-fit Cloudy
models (see text).}
 \label{densidiag}
\end{figure}

In highly ionized plasma the hydrogen number density is related to the electron number density through $n_e \simeq 1.2~ \vy{n}{H}$.
Under the assumption of collisional excitation, the ratio of level populations between \siv* (E=951 cm$^{-1}$) and \siv (E=0 cm$^{-1}$)
provides a direct probe of $n_e$. Using the column densities for \siv\ and \siv* reported in Table~\ref{coldensmo1} and the electron temperature found for our best-fit Cloudy models (10,000 K, a weighted average for \siv\ across the slab),
we find $\log(n_e) = 4.11^{+0.14}_{-0.37}$ cm$^{-3}$ for the outflow in SDSS J1106+1939 (see Figure~\ref{densidiag}). The negative error quoted on the electron number density is computed
from the uncertainties on the \siv\ and
\siv* column densities given in Table~\ref{coldensmo1}. In this case the $\sim 50\%$
statistical error is dominant compared with possible systematic errors
due to our continuum placement, as well as the assumption of a
Gaussian distribution for the absorbing material in the trough.
However, the positive statistical error on the  \siv*\ to  \siv\ ratio
is dominated by the 2.5\% quoted negative error on the \siv\ column
density (see Table~\ref{coldensmo1}). In this case the systematic errors mentioned
above should dominate. In order to derive a more physically plausible and
conservative positive statistical error on the  \siv*\ to  \siv\ ratio
we use the results of our  F1+F2 model (Table~\ref{coldensmo2}). The  different distribution of
the absorbing material for the  F1+F2 model causes a much
larger deviation in the derived \siv* and  \siv\ column density than
possible errors due to our continuum placement. We therefore use the higher
\siv* to \siv\ ratio of the  F1+F2 model as the  positive 1$\sigma$
error for $n_e$.

Using the \siv\ and \siv* column densities reported in Table~\ref{tabj1512_t4}
we find $\log(n_e) \leq 3.34$ cm$^{-3}$ for component 4 of the SDSS J1512+1119
outflow (see Figure~\ref{densidiag}). In Paper II we showed that no information about $n_e$ could be directly derived from the line profile of \siv\ and \siv* for component 2 of the SDSS J1512+1119 outflow. We therefore used the ratio of population between excited states of \ciii* $\lambda 1175$ to constrain
the electronic density and found $\log(n_e) = 5.4^{+2.7}_{-0.6}$ cm$^{-3}$.

\subsubsection{Determining $Q_H$ and $L_{Bol}$}

We compute $Q_H$ as well as the bolometric luminosity $L_{Bol}$ for each object by fitting the UV-soft SED to the measured flux (corrected for
Galactic reddening) at 1100~\AA\ (in the rest-frame) and following the procedure outlined in \citet{Dunn10a}. We obtain $\log{Q_H} =57.1$ s$^{-1}$
and $\log{L_{Bol}} =47.2$ erg s$^{-1}$ for SDSS J1106+1939 and $\log{Q_H} =57.4$ s$^{-1}$ and $\log{L_{Bol}} =47.6$ erg s$^{-1}$ for SDSS J1512+1119.

\subsection{Constraining $\Omega$}
\label{omegadisku}

For a full discussion about constraining $\Omega$ that is needed for
Equations (2) and (3), we refer the reader to Section 5.2
in \citet{Dunn10a}. Here we reproduce the main arguments and then
concentrate on the case of \siv\ outflows. There is no direct way to
obtain the $\Omega$ of a given outflow from
its spectrum as we only see the material located along the line of
sight. Therefore, the common procedure is to use the fraction
of quasars that show outflows as a proxy for  $\Omega$.
Statistically, \civ\  BALs are seen in 20\% of all quasars
\citep[e.g.][]{Hewett03},
and therefore an $\Omega=0.2$ is used for high ionization outflows.
Since the ratio of \civ\ and \siv\ ionic fraction as a function of
$U_H$
is relatively constant (see Paper I) they arise from the same
photoionized plasma. Therefore, we can simply choose $\Omega=0.2$ for
the \siv\
outflows. However, to be more conservative we will multiply the \civ\
$\Omega$ by the fraction of \civ\ BALQSOs that show corresponding
\siv\
troughs. The rationale for this is that, being a less abundant ion than
\civ, \siv\ may only arise in high $N_H$ outflows and therefore a correction
factor to that effect may be needed.

We only compare BAL outflows for two reasons. First, the \siv\ troughs
are located within the Lyman forest.
The spectral coverage of the Sloan Digital Sky Survey (SDSS) can
show these \siv/\siv* troughs only for
redshifts  $z>2.8$, where the forest is very thick and greatly
complicates the identification of narrow \siv\ troughs.
Second, the \civ\ outflows in both SDSS J1106+1939 and SDSS J1512+1119
adhere to the BAL definition \citep{Weymann91}, and therefore the
comparison should be made with BALs only. We note that narrower \civ\
absorption is much more prevalent in quasars, up to
60\% of all objects \citep[see][]{Ganguly08}, and therefore their
$\Omega$ should be larger accordingly.

The sample we showed in Paper I (their Table 1) contains 24 objects
that show appreciable \civ\ absorption.
From these 24 objects, 15 show a \civ\ BAL trough (again using the
\citealt{Weymann91} definition). In Paper I, we identified clear cases of
\siv/\siv* outflows in three of these objects. This was done using the
very restrictive criterion of
matching a \siiv\ absorption template to both \siv\ and \siv*
troughs (see Paper I, Figure 1).
  However, for our purposes here, we need to account for both pure \siv\ outflow
(since low density outflows may not show \siv* even when \siv\
absorption is unambiguous) and also cases
where the \siv\ and \siv* are so wide that they blend into one trough.
We find 3 such strong cases: SDSS J0844+0503
that shows roughly 4000 km s$^{-1}$ continuous \siv/\siv* trough at the
expected velocity;
SDSS J1051+1532 that shows $\sim$ 6000 km s$^{-1}$ continuous \siv/\siv*
trough at the expected velocity;
and SDSS J1503+3641 that shows a good kinematic match for both \siv\
and \siv* based on the \siiv\ absorption template.
There are also 3 plausible cases of matching narrower \siv\ and/or
\siv* absorption features, but we do not include them as
the possibility of a false positive identification of unrelated Ly$\alpha$ absorption features is quite significant. We note that the objects we analyse in this paper are not part of the
the sample presented in paper I; the SDSS spectrum of SDSS J1512+1119 does not cover the \siv/\siv*
troughs region, and SDSS J1106+1939 is below the magnitude cut-off of that survey.

We thus have a total of 6 secure detections of \siv\ outflow troughs
within the 15 \civ\ BAL sample of paper 1 (the three objects shown in
Paper I Figure 1, and the three objects discussed in the paragraph
above). Therefore, a conservative lower limit on the fraction of \civ\ BALQSOs that also
show \siv\ outflows, is 6/15=40\%. We will
use this correction factor for the canonical \civ\ BALQSOs
$\Omega=0.2$, which yields $\Omega=0.08$ for the \siv\ outflows.

\subsection{Results}
\label{res}

Inserting the parameters derived in Sections~\ref{photocon} and ~\ref{firstorder} into Equations (1), (2) and (3), we calculate the
distance and energetics for the three \siv\ outflows. These quantities, as well as most of the parameters they are
derived from, are listed in Table~\ref{tabmflo}. For outflows where ranges of parameters were determined, we show
the corresponding ranges in $R$, $\dot{M}$ and $\dot{E}_k$.

In Table~\ref{tabmflo} we present 3 different models for the outflow of SDSS J1106+1939:
\begin{itemize}
 \item { Using solar abundances and the UV-soft SED (see Section~\ref{photocon} for details). The advantage of this model
is its simplicity and the ease of comparison with any other outflows that are modelled with solar abundances.}
 \item{ Using solar abundances and the canonical MF87 SED developed for AGN by \citet{Mathews87} (see Section~\ref{sensitivity} below
for details),to give a concrete example of the results for a different SED.}
 \item{ A model with metallicity 4 times higher than solar, combined with the UV-soft SED (see Section~\ref{pij1106} for details).
As noted in Section~\ref{pij1106}, we consider this model to be the most physically plausible for SDSS J1106+1939.}
\end{itemize}
The UV-soft SED we use is the most appropriate given the spectral data on hand, and the chemical abundances for $Z= 4 Z_\odot$
give a slightly better fit to the measured column densities than the pure solar case. At the same time, it provides a more
conservative (lower) estimate for the kinetic luminosity of this outflow. We therefore use this model as the representative
result for the outflow.

For the two separate outflow components of SDSS J1512+1119 (C2 and C4), we show only the result of models using
solar abundances and the UV-soft SED.  For comparison, we also show in Table~\ref{tabmflo}
the parameters derived for our previous $\dot{E}_k$
record holder: SDSS J0838+2955 (\citealt{Moe09}, but see the correction by a 0.5 factor reported in \citealt{Edmonds11}).
We note that $\dot{M}$ and $\dot{E}_k$ for SDSS J0838+2955 were derived using $\Omega=0.2$. Therefore, for a fair
comparison with the \siv\ outflows, the SDSS J0838+2955 result should be divided by 2.5 (since for the \siv\ outflows we
use $\Omega=0.08$).



\begin{deluxetable}{lccrrrrrrr}
\rotate
\tablecaption{{\sc Physical properties of the \siv\ quasar outflows.}}
\tablewidth{0pt}
\tablehead{
\colhead{Object}
&\colhead{$\log(L_{Bol})$}
&\colhead{$v$ }
&\colhead{$\log(U_H)$ }
&\colhead{$\log(N_H)$}
&\colhead{$\log(n_e)$}
&\colhead{$R$ }
&\colhead{$\dot{M}$}
&\colhead{$\log(\dot{E}_k)$}
&\colhead{$\dot{E}_k/L_{Bol}$}\\
\colhead{}
&\colhead{(ergs s$^{-1}$)}
&\colhead{(km s$^{-1}$)}
&\colhead{}
&\colhead{($\mathrm{cm}^{-2}$)}
&\colhead{($\mathrm{cm}^{-3}$)}
&\colhead{(kpc)}
&\colhead{(M$_{\odot}$ yr$^{-1}$)}
&\colhead{(ergs s$^{-1}$)}
&\colhead{(\%)}
}

\startdata

J1106+1139$^{\mathrm{a}}$  & 47.2 & -8250 & 0.0$^{+0.2}_{-0.2}$ & 22.8$^{+0.2}_{-0.2}$ & 4.1$^{+0.14}_{-0.37}$  & 0.18$^{+0.11}_{-0.06}$    & 1100$^{+700}_{-400}$  & 46.4$^{+0.2}_{-0.2}$  & 15$^{+10}_{-5}$ \\

J1106+1139$^{\mathrm{a}}$ MF87  & 47.4 & -8250 & -0.2$^{+0.2}_{-0.2}$ & 22.6$^{+0.2}_{-0.2}$ & 4.1$^{+0.14}_{-0.37}$  & 0.29$^{+0.19}_{-0.11}$    & 1100$^{+700}_{-400}$  & 46.4$^{+0.2}_{-0.2}$  & 10$^{+5}_{-3}$ \\

\bf{J1106+1139 $4Z_\odot$} & \bf{47.2} & \bf{-8250} & \bf{-0.5$^{+0.3}_{-0.2}$} & \bf{22.1$^{+0.3}_{-0.1}$} & \bf{4.1$^{+0.14}_{-0.37}$}  & \bf{0.32$^{+0.20}_{-0.14}$}    & \bf{390$^{+300}_{-70}$}  & \bf{46.0$^{+0.3}_{-0.1}$}  & \bf{5$^{+4}_{-1}$} \\

J1512+1119$^{\mathrm{a}}$ (C4) & 47.6 & -1050 & -1.7 to -1.4 & 19.8--20.4 & $\leq$ 3.3    & $>$3.1      & $>$3.4    & $>$42.1  & $\geq$10$^{-4}$ \\

J1512+1119$^{\mathrm{a}}$ (C2) & 47.6 & -1850 & -0.9$^{+0.1}_{-0.1}$ & 21.9$^{+0.1}_{-0.1}$ & 5.4$^{+2.7}_{-0.6}$  &  0.3--0.01$^{\mathrm{b}}$          & 55--1$^{\mathrm{b}}$    & 43.8--42.1$^{\mathrm{b}}$  & $<$ 0.02 \\

\hline
J0838+2955 & 47.5 & $-5000$ & $-2.0^{+0.2}_{-0.2}$& 20.8$^{+0.3}_{-0.3}$ & 3.8 & 3.3$^{+1.5}_{-1.0}$ & 300$^{+210 ~\mathrm{c}}_{-120}$ & 45.4$^{+0.2 ~\mathrm{c}}_{-0.2}$ & 0.8$^{+0.5 ~\mathrm{c}}_{-0.3}$ \\

\enddata

\label{tabmflo}
a) Solar metalicity model.\\
b) The ranges corresponds to the upper and lower error bars on $\vy{n}{e}$, respectively. \\
c) Computed using $\Omega =0.2$ (see text for discussion).
\end{deluxetable}

\section{RELIABILITY OF THE MAIN STEPS IN DETERMINING $\dot{M}$ and $\dot{E}_k$}
\label{cav}

Our reported kinetic luminosity ($\dot{E}_k$) for SDSS J1106+1939 is an order of magnitude larger than the
previous highest value in an established quasar outflow (SDSS J0838+2955, \citealt{Moe09}, see Table~\ref{tabmflo}).
Due to the potential importance of this result to AGN feedback processes, we review in this section the steps
that were taken in order to arrive at this result. Our aim is to address possible caveats or systematic issues
that might affect this result, especially whether $\dot{E}_k$ can differ significantly from the most
plausible value we report in Table~\ref{tabmflo} ($Z=4Z_\odot$ model).

\subsection{Ionic Column Density Extraction and Their Implications}
\label{pointsix}

Reliable measurements of the absorption ionic column densities ($N_{ion}$) in the troughs are crucial for determining
almost every physical aspect of the outflows: ionization equilibrium and abundances, number
density, distance, mass flux, and kinetic luminosity. A firm lower limit on $N_{ion}$ of a given trough is produced by
integrating the apparent optical depth ($\tau_{AOD}$) of the trough across its width \citep[see][]{Borguet12a}. Since
outflow troughs often exhibit non-black saturation, one has to be careful when assessing the actual $N_{ion}$ of a given trough.
As shown in Paper II, the actual $N_{ion}$ can be 1000 times larger than the value inferred from $\tau_{AOD}$. 
This is the reason why we treat any singlet trough measurements (e.g., \siIII~$\lambda 1206$) as well as heavily
blended doublets (e.g., \civ~$\lambda\lambda1548,1551$) as lower limits. For wider separated doublets (in our case the important \pv\ and \siv),
our template fitting allows for a clear distinction between two cases:
a) The fully saturated case where the $\tau_{AOD}$
of both doublet components is identical within the measurement errors. In such a case only a lower limit can be put on
the trough's $N_{ion}$. b) $\tau_{AOD}$ of the blue doublet component is significantly larger than the $\tau_{AOD}$ of
the red doublet component within the measurement errors. In that case, an upper limit for the actual $N_{ion}$ is only a few times larger
than that inferred from $\tau_{AOD}$ and can be measured under the partial covering and/or power-law absorber models
\citep[see][]{Arav05}.

Following these principles, the strength of our 3 most important $N_{ion}$ measurement in the SDSS J1106+1939 outflow is as follows:
\begin{itemize}
 \item \hei*: the two measured troughs show an absorption case that is essentially AOD.  Therefore, the measurement
is very reliable and the small error reflects it's robustness (see Table~\ref{coldensmo1}).
 \item  \pv: In our most conservative case, template fitting shows a $\tau_{AOD}$ ratio of 1:1.1,
which due to the quality of the data is still distinguished from the
fully saturated case of 1:1 ratio. The 1:1.1 ratio yields a significantly higher $N_{ion}$
than the AOD case and also causes the large associated error bars.
 \item \siv: the bottom of the \siv\ trough is black.  Therefore, if our template fitting assumption does not hold  here,
there is a possibility that we severely underestimate the true \siv\ $N_{ion}$.
We note that in such a case the estimated $\dot{E}_k$ will be larger for two compounding reasons: the first being that the $N_{ion}$ ratio of \siv*/\siv\
will become smaller, reducing the deduced $n_e$ of the outflow and hence increasing its distance.  As can be see in Equation (3)
a larger $R$ yields larger $\dot{E}_k$. The second one is that a larger total \siv\ $N_{ion}$ will necessitate a larger $N_H$
in the photoionization solution, which also yields a larger value of $\dot{E}_k$. 
\end{itemize}

\subsection{Photoionization Modelling: Sensitivity to Different SEDs and Abundances}
\label{sensitivity}

The lack of observational constraints on the incident SED, especially in the critical region between
13.6~eV and the soft X-ray region, motivates us to check the sensitivity of the above results to
other AGN SEDs. For comparison we use the MF87 SED developed for AGN by \citet{Mathews87}, which is
both considerably different than the  UV-soft SED used
in our analysis due to it's strong ``blue bump'' and also used extensively in the literature.
 The best-fit photoionization models are parametrized by log~$U_H = -0.2$ and
log~$N_H =22.6$~cm$^{-2}$ for MF87, within 0.2 dex of the values obtained with the UV-soft SED.
We note that, as shown in Table 5, the derived mass flow rate and kinetic luminosity using the MF87 determined $N_H$ and $U_H$
are consistent with  those derived using the UV-soft SED to better than 25\%. This is due to the functional dependencies of these quantities on
$N_H$ and $U_H$, and due to the higher $Q_H$ associated with an MF87 SED (see Equations (1) and (3)).
We conclude that the derived results are only mildly sensitive to other physically plausible quasar SEDs.

In contrast, the derived $\dot{M}$ and $\dot{E}_k$ are more sensitive to departure from solar metallicity. A comparison of
the results in the first and third models given in Table 5 shows the following:
Compared to the $Z=Z_\odot$ model, the  $Z=4Z_\odot$ model has 3 times less $\dot{M}$ and $\dot{E}_k$, while at the same time $R$
is 60\%  larger (due to the lower $U_H$ of the $Z=4Z_\odot$ model).
The photoionization reasons for that behavior are discussed in \citet{Korista08} and \citet{Dunn10a}.
We note that since the super-solar metallicity models gives a better solution for the outflow, and the
upper bound for the metallicity is close to $Z=4Z_\odot$ (see Section~\ref{pij1106}),
our representative model indeed yields a conservative lower limit on $\dot{M}$ and $\dot{E}_k$.

\subsection{Reliability of the Distance Estimate}
\label{R_estimate}

Our distance estimate is derived from measuring the ratio of \siv*/\siv\ column densities. If this ratio is larger than unity
there is a possibility that the \siv* is actually saturated and that the true column density ratio is close to 2.
In this case we can only establish a lower limit on $n_e$ and therefore an upper limit on $R$. The converse is true
for cases where the \siv*/\siv\ column density ratio is smaller than unity, which is the case for our SDSS J1106+1939
measurements. Therefore, the distance we derive for this outflow is technically a lower limit.  We note that using this
$R$ value as a measurement instead of a lower limit yields a lower limit on $\dot{E}_k$ (see Equation (2)). Thus,
again our reported $\dot{E}_k$ is conservative. This point is somewhat different than the argument we presented in
point 3 of Section~\ref{pointsix} above as it holds even if the \siv\ resonance trough is not black.

\subsection{Global Covering Factor ($\Omega$)}
\label{GCF}

As noted in the Introduction, this aspect is one of the major strengths of the current analysis. Based on the usual
statistical approach for \civ\ BALs, the range in $\Omega$ for these \siv\ outflows is between $\Omega$=0.2 (the
value for all high ionization BALs that are represented by \civ\ troughs) and $\Omega$=0.08, which takes into account
that only 40\% of \civ\ BALs show kinematically-corresponding \siv\ absorption (see Section~\ref{omegadisku}). For the
energetics calculation, we chose the more conservative $\Omega$=0.08, which minimizes $\dot{E}_k$.

\subsection{Can The Outflows Hide A Significant Amount of Undetected Mass}
\label{WA_material}

The highest ionization species available in our spectrum is \ovi\
for which we can obtain only a lower limit for $N_{ion}$. This
does not allow us to probe the higher ionization material that is known to exist
in AGN outflows via X-ray observations, the so-called Warm Absorber (WA) material. For example, from Table 3 in \citet{Gabel05b},
the combined column density of the warm absorber in the NGC 3783 outflow is 10--20 times higher than that detected in the UV components.
At least in one case, UV spectra of outflows from luminous quasars show a similar phenomenon
where very high ionization lines (\neviii, and \mgx) show that the bulk of the outflowing material is in this very high ionization component
(see \citealt{Muzahid12}; Arav et al 2012b in preparation). 

It is certainly possible that such a high ionization phase is also present in the SDSS J1106+1939 outflow
(and in the other two \siv\ outflows we report here). In that case, the total $\dot{M}$ and $\dot{E}_k$
could be much larger than reported in Table~\ref{tabmflo} if the WA material is located at the same distance
than the UV outflow.

\subsection{Summary: $10^{46}$ ergs s$^{-1}$ is a conservative lower limit for $\dot{E}_k$ in the SDSS J1106+1939 outflow}

In this section we demonstrated that essentially all possible  deviations from our assumptions will lead to a higher
value of $\dot{E}_k$ in the SDSS J1106+1939 outflow:  possible higher column density in either \siv\ or \pv;
 metallicity lower than $Z=4Z_\odot$; the derived $R$ can only be larger (i.e., if the \siv\ column density is
underestimated); we use a conservative value for $\Omega$, using the one associated with
the general high ionization \civ\ outflows will yield $\dot{E}_k$ larger by a factor of 2.5; the outflow
may very well carry a dominant component of high ionization  material, to which our rest-frame UV observations are
not sensitive; in that case $\dot{E}_k\propto N_H(UV)+N_H(WA)$, where $N_H(UV)$ is the total hydrogen column we are sensitive
to using the UV diagnostics in the data analyzed here, and $N_H(WA)$ is the $N_H$ associated with the higher
ionization WA material.

\section{DISCUSSION}
\label{discufin}

The powerful BAL outflow observed in SDSS J1106+1939 possesses a kinetic luminosity  high enough to play
a major role in AGN feedback processes, which typically require a mechanical energy input of roughly
0.5--5\% of the Eddington luminosity of the quasar \citep[][respectively]{Hopkins10,Scannapieco04}.
This quasar, being bright for its redshift band radiates close to its Eddington limit (i.e. $L_{Bol} \simeq L_{edd}$).
Therefore, with $\dot{E}_k\gtorder 5\%  L_{Bol}$, it has enough energy to drive the theoretically invoked
AGN feedback processes.

How applicable are these results to the majority of quasar outflows?
The investigation described here gives the first reliable estimates of $R$, $\dot{M}$ and $\dot{E}_k$
for a few high ionization, high luminosity quasar outflows. Previously we only had such determinations for low
ionization outflows, which comprise only $\sim$10\% of all  observed quasar outflows. Furthermore, the absorption spectrum of these
two objects looks very similar to the run-of-the-mill BALQSO spectra longward of \Lya. This is an important point
as the vast majority of available BALQSO spectra do
not extend to  wavelengths shorter than 1100\AA\ (restframe) and therefore do not cover the \siv/\siv* lines. 
The phenomenological similarity of the \civ\ and \siiv\ BALs in the objects presented here to the majority of
observed BALs suggests that a straightforward generalization of the results may be plausible. SDSS J1106+1939
is the first VLT follow-up observation we obtained
from a dedicated search for possible \siv/\siv* troughs. In the coming year, we are scheduled to obtain several more
X-shooter observations  of such candidates and will be able to shed more light on this issue.

The distances found for the three outflows we report here
range from $\sim100$ parsecs to a few kiloparsecs. These are similar to the distances inferred for outflows in which
the density diagnostic is obtained from the study of excited troughs of singly ionized species as \feii\ or \siII\
\citep[e.g.][]{Korista08,Moe09,Dunn10a}, but they are 3+ orders of magnitude further away than the assumed acceleration region
(0.03--0.1 pc) of line driven winds \citep[e.g.][]{Murray95,Proga00}. This result is consistent with
almost all the distances reported for AGN outflows in the literature. However, the current research expands the
claim to the majority of high ionization outflows.  We conclude that most AGN outflows are observed very
far from their initial assumed acceleration region.

\section*{ACKNOWLEDGMENTS}

B.B. would like to thank Pat Hall for suggesting the use of the revised redshifts. We acknowledge support from NASA STScI grants GO 11686 and GO
12022 as well as NSF grant AST 0837880.


\bibliographystyle{apj}

\bibliography{astro}{}


\end{document}